\documentclass[a4paper,10pt]{article}
\usepackage[utf8]{inputenc}
\usepackage{pmboxdraw}
\usepackage{amsmath}
\usepackage{amsthm}
\usepackage{amssymb}
\usepackage{pgfplots}
\usepackage{subcaption}
\pgfplotsset{compat=newest}
\usepackage{tikz}
\usepackage{svg}
\usepackage[shortlabels]{enumitem}
\usepackage{tikz-cd}
\usepackage{faktor}
\usepackage{mathtools}
\usepackage{geometry}
\usepackage{authblk}
\usepackage{listings}
\usepackage{changepage}
\usepackage{xcolor}
\usepackage{csquotes}
\usepackage{natbib}
\usepackage[hidelinks]{hyperref}
\hypersetup{colorlinks=true, urlcolor=blue, linkcolor=teal, filecolor=purple, citecolor=teal}

\theoremstyle{definition}

\newtheorem*{example*}{Example}

\newtheorem*{problem}{Problem}

\newcommand{\geosolver}{Newclid}

\makeatletter
\DeclareRobustCommand\vttfamily{%
  \not@math@alphabet\vttfamily\relax
  \fontfamily{cmvtt} %
  \selectfont
}
\DeclareTextFontCommand{\textvtt}{\vttfamily}

\makeatother

\setlist[itemize]{align=parleft,left=0pt..1em}
\colorlet{numb}{magenta!80!black}
\definecolor{darkcolor}{RGB}{127,0,85}
\lstdefinelanguage{jgex}{
    basicstyle=\fontsize{8.6}{11}\vttfamily\color{blue},
    commentstyle=\color{black},
    stringstyle=\color{darkcolor},
    showstringspaces=false,
    breaklines=true,
    breakatwhitespace=true,
    frame=lines,
    string=[s]{"}{"},
    comment=[l]{:\ "},
    morecomment=[l]{:"},
    literate=
        *{=}{{{\color{numb}=}}}{1}
        {;}{{{\color{numb};}}}{1}
        {?}{{{\color{numb}?}}}{1}
        {,}{{{\color{numb},}}}{1}
        {⟂}{{$\perp$}}{1} %
        {∠}{{$\angle$}}{1} %
        {⇒}{{$\Rightarrow$}}{1} %
        {∥}{{$\parallel$}}{1} %
}

\newcommand*\samethanks[1][\value{footnote}]{\footnotemark[#1]}

\title{\geosolver: A User-Friendly Replacement for AlphaGeometry}

\author[1]{Vladmir Sicca\thanks{Equal contributions.}%
\thanks{Corresponding author: \texttt{vladmir.siccagoncalves@mail.mcgill.ca}}}
\author[3]{Tianxiang Xia\samethanks[1]\thanks{Work carried out while at Huawei.}}
\author[ ]{Math\"is F\'ed\'erico\samethanks[1]\samethanks[3]}
\author[4]{Philip John Gorinski\samethanks}
\author[2]{Simon Frieder\thanks{Corresponding senior author: \texttt{simon.frieder@cs.ox.ac.uk}}}
\author[1]{Shangling Jui}
\affil[1]{Huawei Lagrange Mathematics and Computing Research Center}    
\affil[2]{Oxford University}
\affil[3]{ETH Zürich}
\affil[4]{Robin AI}

\begin{document}
\maketitle

\begin{abstract}
We introduce a new symbolic solver for geometry, called \geosolver, which is based on AlphaGeometry.
\geosolver{} contains a symbolic solver called DDARN (derived from DDAR-\geosolver), which is a significant refactoring and upgrade of AlphaGeometry's DDAR symbolic solver by being more user-friendly - both for the end user as well as for a programmer wishing to extend the codebase. For the programmer, improvements include a modularized codebase and new debugging and visualization tools. For the user, \geosolver{} contains a new command line interface (CLI) that provides interfaces for agents to guide DDARN. DDARN is flexible with respect to its internal reasoning, which can be steered by agents. Further, we support input from GeoGebra to make \geosolver{} accessible for educational contexts.
Further, the scope of problems that \geosolver{} can solve has been expanded to include the ability to have an improved understanding of metric geometry concepts (length, angle) and to use theorems such as the Pythagorean theorem in proofs. Bugs have been fixed, and reproducibility has been improved. Lastly, we re-evaluated the five remaining problems from the original AG-30 dataset that AlphaGeometry was not able to solve and contrasted them with the abilities of DDARN, running in breadth-first-search agentic mode (which corresponds to how DDARN runs by default), finding that DDARN solves an additional problem. We have open-sourced our code under:
\begin{center}
    \url{https://github.com/LMCRC/Newclid}
\end{center}

\end{abstract}

\section{Introduction}
\label{sec:introduction}

\paragraph{General remarks.} AlphaGeometry~\citep{trinh_solving_2024} demonstrated the ability to solve geometry problems at the level of the International Mathematical Olympiad (IMO), with performance comparable to top human competitors. At the heart of AlphaGeometry is a formal language that encodes geometric problems and theorems, rooted in JGEX~\citep{ye2011introduction}, as well as a symbolic reasoning engine called DDAR (see Subsection~\ref{subsec:modifying_ar} for more information), written in Python, which is an extension and reimplementation by~\citet{trinh_solving_2024} of an earlier symbolic AI engine~\citep{chou2000deductive}. 

Intertwined with DDAR in the original work is a large language model (LLM), trained on a synthetic dataset of proofs generated using DDAR, that predicts new geometric clauses that DDAR can use to draw new inferences.

\paragraph{AlphaGeometry's inference loop.} AlphaGeometry works, at a high level, in the following way: DDAR iterates through statements that can be logically justified from previous ones until it finds what needs to be proved. There are many ways to find a new statement: An initial batch of them comes from expanding the clauses into statements by the rules described in the \texttt{defs.txt} file. Then, sequential iterations of the AR module that make up DDAR (see Subsection \ref{subsec:modifying_ar} for more information), followed by the application of rules in the \texttt{rules.txt} file is applied until the goal or a fixpoint is reached. In case of reaching a fixpoint, the LLM will be called to add a new clause so the iterations can start again. 

\paragraph{Issues.} Despite its impressive reasoning capabilities, AlphaGeometry suffers from limitations in terms of user-friendliness, both for the end-user as well as for the programmer interested in extending the current codebase and the scope of problems it can handle.

\begin{itemize}
\item \textbf{User-friendliness issues:} There are three main obstacles that users have to overcome to use AlphaGeometry: 
\begin{itemize}
    \item Installing AlphaGeometry is difficult, mainly because the Meliad library,\footnote{\url{https://github.com/google-research/meliad}} on which AlphaGeometry depends, is difficult to install.
    \item Problems have to be input using the JGEX formal language.
    \item The files \texttt{rules.txt} and \texttt{defs.txt} inside the AlphaGeometry system describe the foundations which DDAR uses to make inferences.\footnote{Detailed descriptions and explanations of the content of those files can be found in the project's documentation under \url{https://lmcrc.github.io/Newclid/}}
    Yet, during AlphaGeometry's inference loop (see above), different rules could kick that were not explicitly stated in these text files, but are hard-coded, see Section~\ref{subsec:adding_new_rules}.
\end{itemize}

\item \textbf{Coder-friendliness issues.}
AlphaGeometry's codebase is not modularly built, which makes it hard for someone who want to contribute code to add new features, inspect proof traces, add logs, etc. Further, the LLM is implemented in Meliad, a deep-learning Python library that is not widely used - which further makes it hard to finetune the LLM or understand its inner workings, among other desirable operations.

\item \textbf{Problem scope issues.}
AlphaGeometry is not able to work with rather simple and fundamental theorems, notably the Pythagorean theorem, lacking support for the concept of length of a segment, used, among other things, in elementary geometry courses. 
This makes AlphaGeometry a type of {\textit{narrow}} AI system whose intelligence contrasts with human intelligence: it is inconceivable that an IMO-level competitor will be able to solve certain IMO-level geometry problems while not being able to use the Pythagorean theorem. AlphaGeometry \enquote{overfits} on Olympiad geometry problems, compared to arbitrary plane geometry.
\end{itemize}

\paragraph{Contributions.} In this research work, we have focused our efforts predominantly on the DDAR solver (as opposed to the LLM) because of the relative importance of DDAR when compared to the LLM in considering the contributions of these two core components of AlphaGeometry towards its final performance on the test dataset: Two datasets were used to report the performance of AlphaGeometry: AG-30, consisting of 30 geometry problems from the IMO, as well as a larger unnamed set of 231 Olympiad-level geometry problems~\citep{trinh_solving_2024} (no data collection protocol has been provided). We will call the latter \emph{AG-231} to distinguish between the two. Table 1 in~\citep{trinh_solving_2024} presents a breakdown of the performance of AlphaGeometry, as a function of which combination of methods has been used (DD only, DDAR, DDAR + human heuristic, DDAR + various forms of the LLM, etc.) on the AG-30 dataset. This is augmented by Figure 6,b in the Extended Data section of~\citep{trinh_solving_2024}: If we re-express the results on AG-30 from Table 1 in percentages, the best symbolic AI approach (DDAR + human heuristics) solves 60\% of the problems, whereas the best deep-learning approach (DDAR + LLM) solves $\sim$83.3\% of the problems. Yet this increase of $\sim$23\% is reduced to an increase of merely $\sim$6.5\% on the larger---and thus more representative---AG-231 dataset, where DDAR + human heuristics solve $\sim$92.2\%, and DDAR + LLM solves $\sim$98.7\% of problems. 
(Further, it was noted in~\citep{sinha2024wu} that the number reported in Table 1 on how many problems could be solved by the classical Wu method~\citep{wu1978decision} was underestimated since it was found that Wu's method was able to solve 15 rather than 10 problems.)

Concretely, because of DDAR's outsized importance in AlphaGeometry, as argued above, we focused on \geosolver{}, which uses a new symbolic solver called \enquote{DDARN} (DDAR-\geosolver) that fixed many of the issues present in DDAR:

\begin{enumerate}

\item \textbf{User friendliness.} The following elements were improved:

\begin{itemize}
    \item Simplified installation: We provide simple ways to install \geosolver, in particular, via PyPI. We have removed the dependency on the Meliad library, and streamlined the installation process, see Subsection~\ref{subsec:easy_install}.
    \item Problems can be input using an improved command-line interface (CLI), which offers endpoints to introduce agents that can manipulate DDARN, see Subsection~\ref{subsec:cli}.
    \item Additionally to the new CLI, problems can also be input using the GeoGebra interface, see Subsection~\ref{sec:geogebra_interface}.
    \item We have slightly expanded the scope of problems that can be solved. DDARN can use, in particular, Pythagoras theorem, see Subsection~\ref{subsec:problem_scope}.
\end{itemize}

\item \textbf{Coder friendliness.}  A significant refactoring of existing classes was performed, and several tools that assist in debugging and allow the visualizing of several internal objects (such as symbolic graphs) are introduced. These improvements fall into three categories: 

\begin{itemize}
    \item General code refactoring that does not affect the reasoning capabilities but laid the groundwork for any subsequent work that was done, see Subsection~\ref{subsec:foundational_design} 
    \item improvements that affect how the reasoning works, which include changes that make adding future code easier, see Subsections~\ref{subsec:agentic}, and
    \item improvements that improve the DDARN reasoning engine, see Subsection~\ref{subsec:engine_improve}, and
    \item tools that make debugging and visualizing of DDARN's internal objects easier (in fact we used these to find some of the missing rules, see Subsection~\ref{subsec:debugging}.

\end{itemize}

\item \textbf{Reproducibility.} We have elaborated on the reproducibility of AlphaGeometry and \geosolver{}, see Section~\ref{sec:reproducibility}.

\item  \textbf{Detailed evaluation.} Lastly, we compare AlphaGeometry's DDAR to  \geosolver's DDARN on the five problems from the AG-30 dataset whose solution eluded AlphaGeometry, namely the IMO problems 2008 P1B, 2006 P6, 2011 P6, 2019 P2 and 2021 P3. Two problems out of these five, we argue, cannot easily be solved given the current symbolic solver framework; one problem that previously was not solvable becomes solvable using DDARN (IMO 2008 P1B), as can be seen in Section~\ref{sec:missing_problems}. 

\end{enumerate}

In the future (see Section~\ref{sec:user_features}), we plan to augment \geosolver{} to also include an improved LLM, compared to the one present in AlphaGeometry, that would be one example of an agent that manipulates DDARN.

\section{Terminology} 
\label{sec:terminology}
Below, we define various terms related to functional aspects that underlie both DDAR and DDARN. Some (but not all) of these were implicitly used in~\citep{trinh_solving_2024}. To make the inner workings of DDAR and DDARN reasoning engines more transparent, we chose to give an explicit description of all of them to enhance the conceptual understanding of DDAR and DDARN.

When defining a theory for plane geometry, one is faced with the question of deciding what are the fundamental objects that will be used to describe the theory. The choice made for DDAR, and continued in DDARN, are to base the theory on points, and use them to represent other geometric objects (lines, circles, triangles, angles, ratios, relationships between objects, etc.).
The relationship descriptors, such as \texttt{cyclic}, or \texttt{cong}, to denote whether a collection of points lie on a circle or two congruent segments, respectively, are called \emph{predicates}. Their use dates back to the DD symbolic engine introduced by~\citet{chou2000deductive}. Predicates work thus like a function, taking as arguments points, e.g., \texttt{cyclic a b c d} describes that the four points \texttt{a, b, c, d} are on the same circle and \texttt{cong a b c d} describes the congruence of the two segments, {texttt{ab} and \texttt{cd}, that are made up by the four points \texttt{a,b,c,d}. A predicate that is instantiated by points, such as the mentioned \texttt{circle a b c}, is called a \emph{statement} and is the foundational element of the reasoning engine.

In DDAR and DDARN, inferences can be made by using three different paths: application of rules described in the \texttt{rules.txt} file, running the algebraic reasoning module, and resolutions that are made on the go through hard-built, not-described functions (which we call \emph{intrinsic rules}).

The collection of statements that DDAR and DDARN store at a given point when solving a problem is the {\emph{proof state}} of the problem at that stage. The objective of the engine is to find a specific statement, the {\emph{goal}}, supplied by the user, in the proof state. 

Whenever we use the word {\emph{symbolically}} in this article, we refer to something that is inferred exclusively from the proof state.
As explained below, statements are not directly inserted in the proof state by a human but are either derived from the problem prescription, derived from previous statements by the inference loop that runs within DDAR and DDARN, or, in some rare cases, directly derived from a diagram of the problem through a numerical check. 

The problem prescription, created by a human user, does not use the language of the statements and predicates apart from the establishing of the goal, but that of {\emph{clauses}}, implemented from {\emph{definitions}}. Clauses represent geometric constructions, and we will often refer to clauses as \emph{constructions}. Most clauses will induce the generation of statements. For example, in a problem where points \texttt{a}, \texttt{b}, and \texttt{c} are already defined, we may define point \texttt{d = incenter d a b c} in the statement of the problem. 
In that case, when the problem is read, the statements described in the \texttt{incenter} definition, with those arguments, will be added to the starting point of the proof state, namely the ones corresponding to the fact that the incenter is the meeting of the bisectors of the angles of the triangle: \texttt{eqangle a b a c a x a c}, \texttt{eqangle c a c x c b}, and \texttt{eqangle b c b x b x b a}.

The definition \texttt{incenter} will also provide instructions to get numerical coordinates for point \texttt{d}, and here the information that the incenter is the intersection of internal bisectors of the angles of the triangle instead of external ones is recorded in some sense. In comparison, there is also a definition of \texttt{excenter} that is symbolically equivalent to the \texttt{incenter}, which will create a point with an appropriately different numerical representation. Clauses can also be added by the LLM later in the solving loop of a problem, but technically DDAR/DDARN see this action as the writing of a new problem.

During the reasoning procedure, the main way to generate new statements is through the applications of {\textit{rules}}. A rule is the encoding of a theorem, and it consists of a sequence of statements,
the hypothesis, followed by a new statement, the consequence. Of course, a rule is written with generic points, so a separate procedure when rules are actually being used is to do the {\textit{matching}} of the rule. The matching is the process of seeing if, for some combination of points, all the hypotheses of the rule are statements in the proof state. If so, the consequence of the rule will be added to the proof state as well. A full list of rules available for \geosolver{}, numbered \texttt{r1} to \texttt{r69}, can be found in Appendix~\ref{app:rules}.

These shortcomings prevent it from being widely adopted by mathematicians working in the domain of theorem proving, educators, and students learning Euclidean geometry. Therefore we introduce \geosolver, which is a large refactoring of AlphaGeometry, that improves AlphaGeometry's codebase, in particular the DDAR solver.

\section{User Friendliness.}

Below, we detail the four improvements that we made regarding user-friendliness: This includes a simplified installation process, two interfaces that we have added to \geosolver{} to make it more user-friendly (the command line interface (CLI), and the other is a GeoGebra interface), and a section on how we expanded the problem scope that DDARN can solve.

\subsection{Easy Installation}
\label{subsec:easy_install}

We have streamlined the installation process so that it is easy to install \geosolver{} using \texttt{pip install}; additionally, we are releasing it as a PyPI package. The isolation of the Meliad dependency was the main factor in making the installation easier, as running the DDARN solver now is not contingent on having Meliad installed (for AlphaGeometry, even if only DDAR was used, Meliad also had to be installed). 
Further, we have created an API so that \geosolver{} can be comfortably called from other code.

\subsection{Command Line Interface (CLI)}
\label{subsec:cli}

Our most fundamental new interface is presented to the user via the command line interface (CLI). Its main objective is to allow one to run a problem with \geosolver, without having to use a Python code entry point, and with the option of a human-understandable step-by-step process, which allows the replacement of the original LLM by human decisions.

The CLI is characterized by a high degree of control over the solver through our current three agents (a human option, a brute-force automatic option, or a dummy option), which we detail below.

Problems always need a defined name: multiple problems can be inside a single text file and are referred to by a name. In the case of the GeoGebra folder with the GeoGebra \texttt{.ggb} file (which contains the entire construction made up of clauses but not the goal), a text file with the goal needs to be supplied.
Problems can be fed to \geosolver{} in formal language from a \texttt{.txt} file, just as in the case of AlphaGeometry. The CLI provides the following elements:
 \begin{itemize}
 \item An environment: We have introduced a new concept of \emph{environment} that contains the rules file and the definitions file for the problems used by the reasoning agent.

 A typical environment, with the outputs of solutions of a hypotetical \texttt{problem1} has the following structure and contents.
 
 \begin{verbatim}
 environment/
 |-- rules.txt
 |-- defs.txt
 |-- problem1/
     | -- problem1.ggb
     | -- html
          | -- index.html
          | -- symbols_graph.html
          | -- dependency_graph.html
          | -- figure.svg
     | -- goals.txt
     | -- construction_figure.svg
     | -- run_infos.txt
     | -- proof_steps.txt
     | -- proof_figure.txt
 |-- problem2/
 \end{verbatim}
 
 Different files with rules and definitions can be used for each problem so that the solver uses a different set of those exclusively for one specific problem. For technical details we refer to the documentation section.

Introducing the concept of environment for a problem proved to be a good solution to 1) flexibly allow various ways of inputs (currently \geosolver{} supports CLI and GeoGebra (see below), but we invite the community to extend our approach), where the environment collects all necessary to enable that input, and 2) to provide, similarly to a software package manager, such as Python's virtual environments, an easy way to run different mathematical problems with different inputs.

 \item Seeds: The construction of a numerical diagram of the problem contains various random choices, which can influence numerical check predicates and change poofs. A seed is collected so that these choices  can be made deterministic;
 \item Display settings and logging: It is specified what logs the terminal should show if the matrices generated by the algebra reasoning sub-engine (the \enquote{AR} part) should be displayed or if no logs should be recorded.
 \end{itemize} 

AlphaGeometry's design is such that it runs, given an initial human translation of a geometric problem into JGEX, to end without further human intervention. Thus, the reasoning procedure is rigid.

Contrasting, we designed \geosolver{} by generalizing its working to allow agentic interaction modes. In particular, the Human Agent is an interface that allows a person to control the path of a solution, deciding which theorems to try to apply at each moment, add new points on the go, check symbolically if a given statement was added to the proof state at any given stage, look at the evolution of the proof graphs at each step, or run the usual breadth-first-search DDAR reasoning procedure on call.

The Human Agent is called as a reasoning agent, currently in opposition to DDARN (which we also treat as an agent, see Subsection~\ref{subsec:agentic}  to see how this is treated in the code), as it can be viewed as a particular way to guide the search for a proof of the problem. It was designed as a first step towards a plan to try to improve the proof search from BFS to a better heuristic through learning from human data of trying to solve problems, but it also proved an important tool in debugging, as it became a way to check the effect of theorems one at a time, drastically increasing the granularity of diagnostics when compared to full BFS runs.

It is still limited for human use because the lack of a better graphic interface can make it hard to directly understand the problems and theorems only from the text, but it is our belief that a friendlier interface for the Human Agent could even make it an interesting learning tool. With it, users can try to experiment fast when solving a problem, without the risk of making false claims, and check what middle steps have been already verified at each point, as well as see the history of the proof state. It could work for problem investigation in plane geometry in tandem with a graphic tool like GeoGebra, the same way symbolic computing and graphic software like Mathematica and MatLab work with investigations in other areas of mathematics and applied sciences.

 \begin{figure}
     \centering
     \includegraphics[width=12cm]{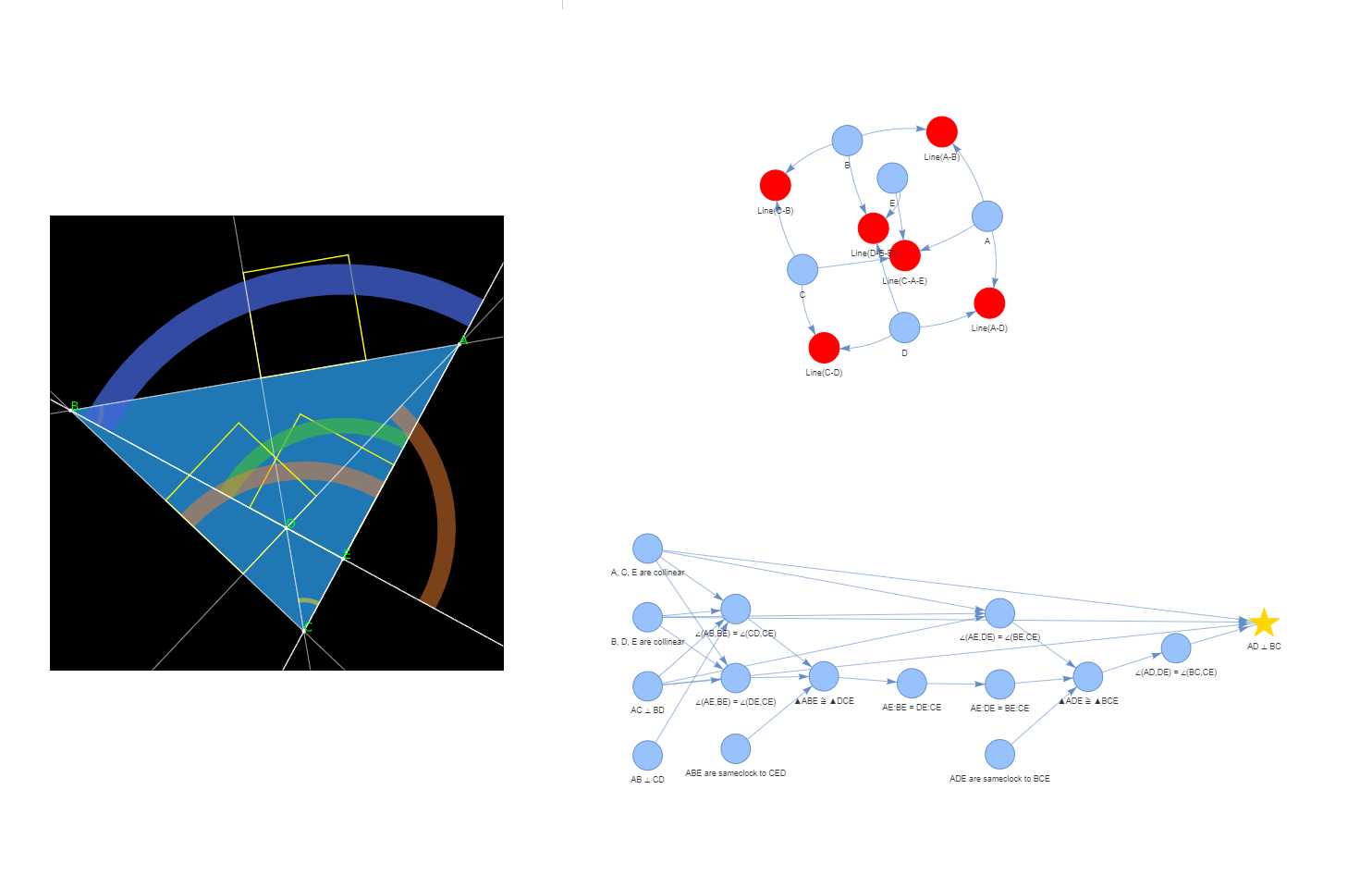}
     \caption{Index page generated for a single problem, with the diagram of the proof, the symbols graph, and the dependency graph.}
 \end{figure}

\subsection{Geogebra Interface}
\label{sec:geogebra_interface}

One barrier to using AlphaGeometry is the need to translate a given geometric problem into the internal JGEX formal language format, which is not widely known, not widely used in other systems, and not well-documented. To make usage of this software simpler, \geosolver{} provides an interface for a user to be able to prescribe the problem from a GeoGebra diagram of the problem instead. The big advantage is that GeoGebra has an intuitive graphic interface, is widely known by people working with Euclidean geometry (including in educational settings), and has a large community to support one's path to learning its usage. 

To provide the statement of the problem, one has to provide a file environment containing the \texttt{.ggb} file and a \texttt{goals.txt} file containing the goals in JGEX format. \geosolver{} will then use the GeoGebra construction to generate the numerical representation of the problem and the construction tools used in GeoGebra to enumerate the initial premises in the proof state, allowing the solver to operate as if the problem were prescribed in text.

\begin{figure}
    \centering
    \includegraphics[width=\linewidth]{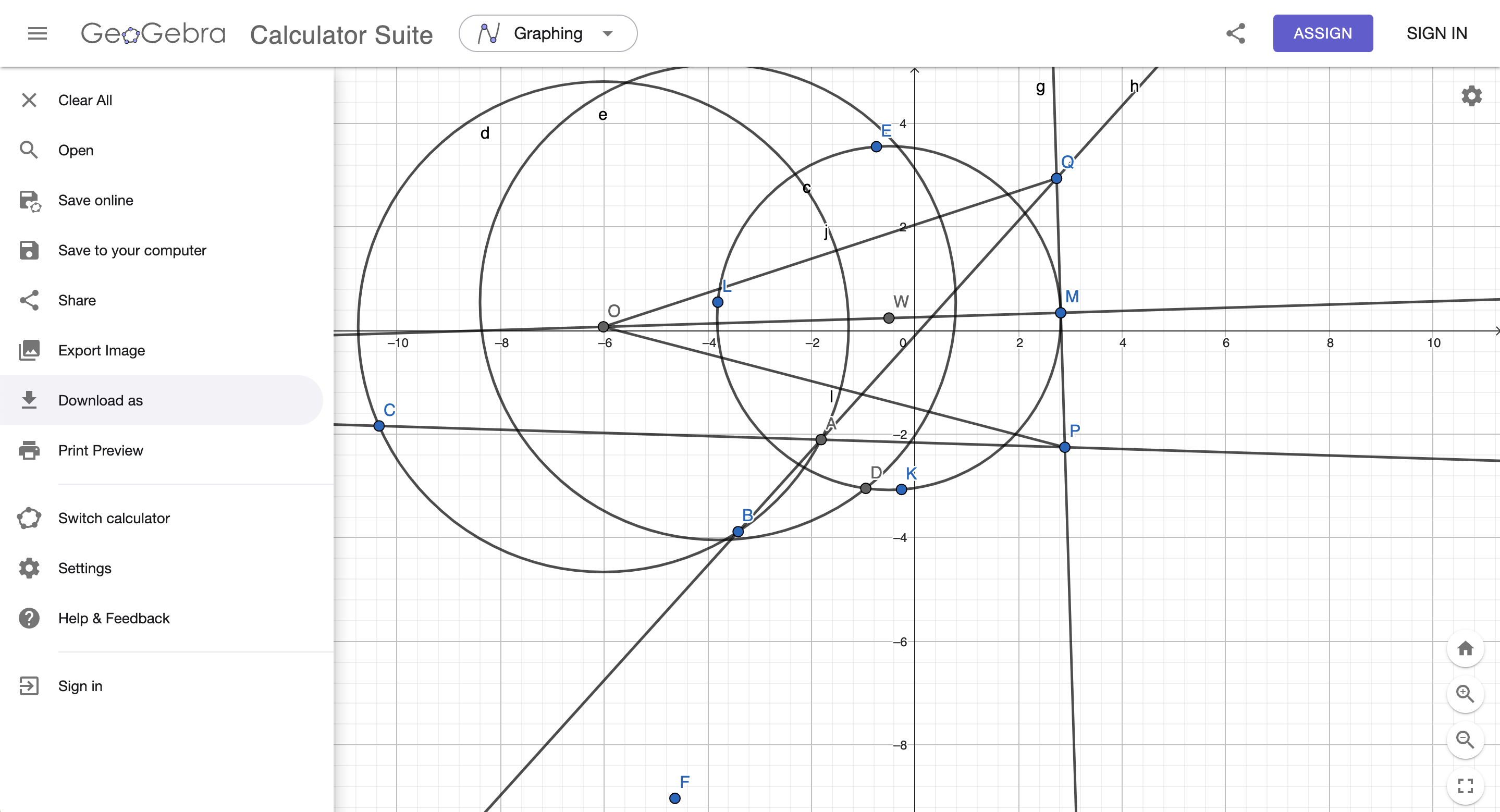}
    \includegraphics[width=\linewidth]{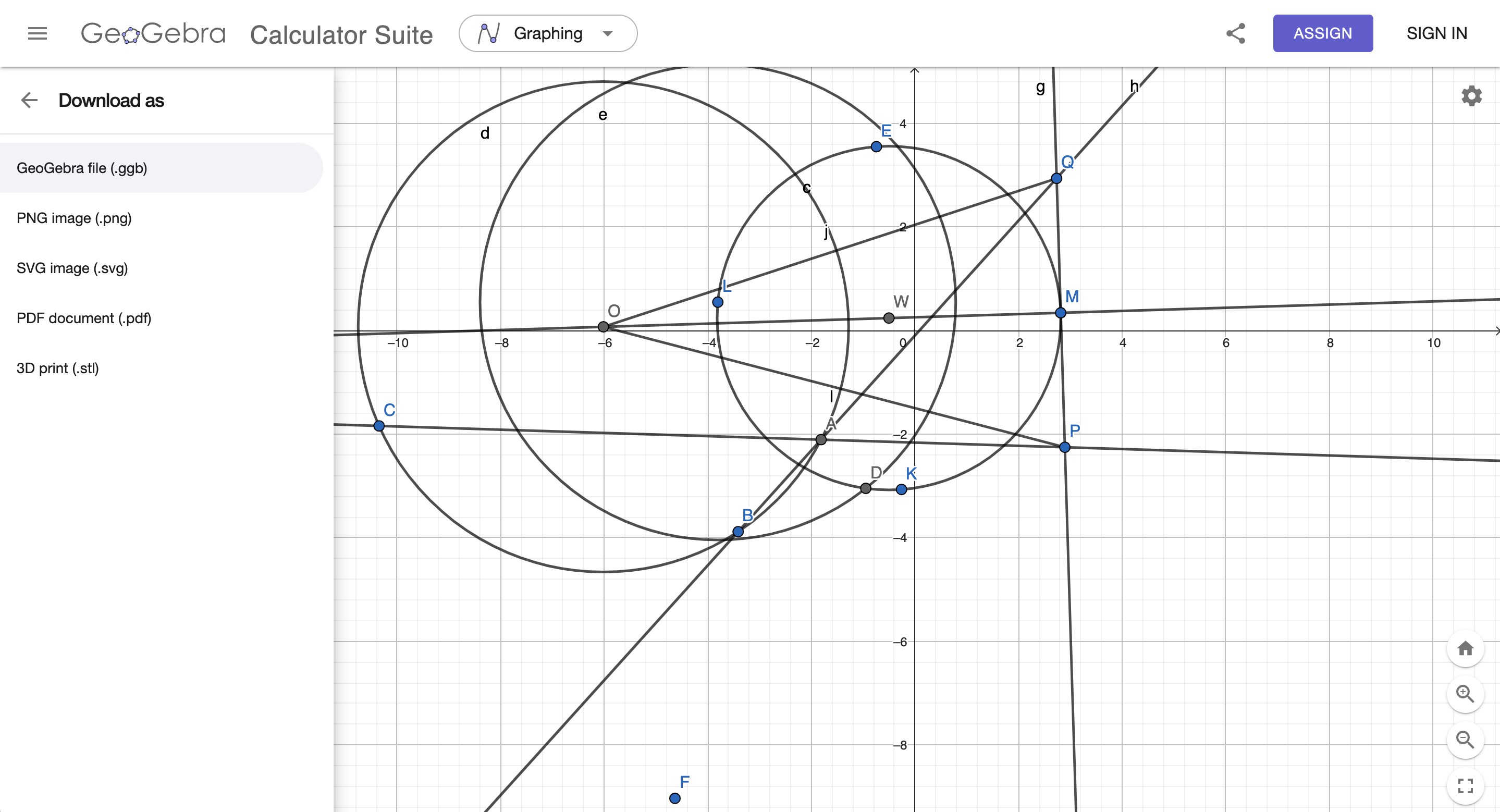}
    \caption{An exemplary problem in GeoGebra that can be parsed by \geosolver.}
    \label{fig:geogebra}
\end{figure}

\subsection{Expanded Problem Scope}
\label{subsec:problem_scope}

All of the extensions outlined here rely on our significant refactoring of the code, which was the foundation that made the extensions possible in terms of reasoning capabilities. In the following Subsections~\ref{subsec:adding_new_predicates},~\ref{subsec:adding_new_definitions},~\ref{subsec:adding_new_rules},~\ref{subsec:adding_ar_equations}, we mention various new additions we made. These are the by-product of our refactoring, which made these extensions, in particular, the ones made in the last section,~\ref{subsec:adding_ar_equations}, possible. Our big achievement is that \geosolver{} is able to use the Pythagorean theorem inside other proofs, see Subsection~\ref{subsec:adding_new_rules}.

\subsubsection{Adding New Predicates}
\label{subsec:adding_new_predicates}

Predicates form the internal vocabulary used by the reasoning of the engine. We recall from the section on terminology, Section~\ref{sec:terminology}, that each predicate behaves as a function, with a prescribed number of arguments which correspond to points. A predicate should be thought by the user as a fact relating the arguments, such as \enquote{the segments defined by two pairs of points being congruent}, or \enquote{the lines through these two pairs of points being parallel}. The only human inputs of predicates are the goals of a problem, but any definition will be symbolically broken down into instances of predicates, and rules are stated in terms of predicates as well. The engine then proceeds to \enquote{think} using this language. Therefore, the list of predicates is the most fundamental list of notions that the engine knows.

Crucially, AlphaGeometry lacked functional predicates that consider numerical measures of angles, ratios, and lengths. This made simple questions, such as finding the third angle of a triangle given the values of the first two, unanswerable, even if the algebraic module could easily find the information. Given that questions asking for a specific angle or distance are extremely common in Euclidean geometry and would be expected of any plane geometry solver, this was a significant limitation we sought to overcome.
(Curiously, we found traces of such predicates inside the released codebase, which left us with the impression that the original intention was to include them but that this was abandoned for reasons unknown to us.)

To fill this gap, we patched up or added the following predicates:

\begin{itemize}
    \item \textbf{aconst.} This predicate encodes the information that the angle between two lines has a given constant value (that can be given in degrees or in radians). Specifically, lines are denoted in order, each by a pair of points, and the angle is the one obtained by going in the counterclockwise direction from the first to the second line. In the original work, this predicate backed the \texttt{s\_angle} definition and superficial predicate, but it only accepted radians as input and could not be used for goals. We fixed it for all uses, created syntax for both degrees and radians in a unified predicate and turned \texttt{s\_angle} into only a definition that directly calls \texttt{aconst}.
    \item \textbf{rconst.} This predicate encodes the information that the ratio between two segments is a given numerical fraction. It was present in the original codebase, in the definition \texttt{triangle12} specifically, but was not functional. We fixed it in all functionalities and switched to a fraction notation for the ratio instead of a pair of integers.
    \item \textbf{lconst.} This predicate encodes the information that the length of a segment is a given numerical value (without units). It was created from scratch, as it was not a functionality within the original AlphaGeometry, so we added it mirroring the behaviour of the other predicates dealing with constant measures. To give consistency to this predicate, it was also necessary to change some of the existing definitions, as their corresponding numerical sketch was not length-agnostic. This complication was probably not addressed in the original software because its whole logic structure was scale invariant by not ever assigning numerical length information. Theoretically, the study of lengths creates a new issue, as a numerical length demands a unit of measurement in contrast to angles and ratios. Our approach to this was to assume a single unnamed unit of measurement throughout each problem.
    \item \textbf{acompute.} This is a new kind of predicate that does not simply ask for the proof of a goal but creates the goal itself on the run. It is meant for open-ended questions, specifically when a problem asks the measure of a specific angle. It will try to locate the corresponding \texttt{aconst} statement and turn it into the goal internally. Then, it generates a proof in which the last statement is the angle being measured equals its value.
    \item \textbf{rcompute.} The exact analogue of \texttt{acompute} for problems asking the value of a ratio between two segments. Similarly, it will search for an \texttt{rconst} statement in the proof state corresponding to the ratio asked and turn it into the goal of the problem. 
    \item \textbf{lcompute.} Working just as \texttt{acompute} and \texttt{rcompute}, \texttt{lcompute} is meant to allow one to ask the length of a segment by searching the appropriate \texttt{lconst} statement. These three predicates add little in terms of logic but are ubiquitous among elementary school problems and allow one to \enquote{solve} geometry problems in a larger sense: not only writing down proofs but also finding open solutions.
    \item \textbf{nsameside.} This predicate is of a different nature with respect to the previous ones, as it is simply a numerical check, not verifiable symbolically. It can be imposed as a goal, but if it is true, it will automatically be satisfied. Adding predicates like this is important to allow for proper enunciating of rules, especially since the solver uses the notion of full angles, which do not differentiate between the two angles in an intersection. The previous version of AlphaGeometry already had a \texttt{sameside} predicate that verified if two points were on the same side of two corresponding vectors by checking if the inner products between pairs of vectors in the configuration have the same sign. For a larger flexibility of adding rules, we added a predicate for the negation of \texttt{sameside} as well, just as was the case for the collinearity, perpendicularity, and parallelism predicates.
    \item \textbf{same\_clock.} Just as was the case for \texttt{nsameside}, \texttt{same\_clock} is also simply a numerical check, one that verifies if the triplets of vertices of two different triangles are ordered with the same orientation. This is important when checking the similarity and congruence of triangles in the case of full angles, as the fact that two angles are the same only if there is an orientation-preserving isometry taking one to the other (with markings) means the rules for triangle similarity and congruence are also orientation sensitive. In the original work, the authors solved this problem by implementing a function that checked orientation hidden inside the matching of rules involving congruence and similarity. The issue, of course, is that such checks were hidden from the proofs and were not highlighted in the code. To solve those problems, we turned the original \texttt{same\_clock} Python function into an explicit predicate that can be used to write rules and that behaves as the other numerical checks.
\end{itemize}

\subsubsection{Adding New Definitions}
\label{subsec:adding_new_definitions}

While predicates are prerequisites for an extension of reasoning, simply adding a predicate does not incorporate it into the engine reasoning. Information about the predicate enters the reasoning loop either by direct insertion of information in the statement of the problem or as a derivation of a rule. As can be noted from the experience of solving geometry problems in general, for problems where angles and lengths measures occur, it is usually necessary that some previous measure is provided, such as a known angle. That is particularly true in the case of lengths, as establishing a scale is always necessary in order to calculate distances.

In parallel, even predicates that are often derived from rules without a {\emph{need}} for an introduction of the statement can be wanted as a premise. For example, the \texttt{eqratio} statement, which corresponds to the fact that the ratios between two pairs of segments are the same, usually occurs as a consequence of verification of conditions on similar triangles, but it is not unusual that a problem has such as a statement mentioned in its premises, in the form of a proportion.

To be able to insert those sorts of premises for problems, \geosolver{} has to introduce new definitions. Introducing new definitions on the syntax is not a hard task from a technical viewpoint, in essence one only needs to define the characteristics of the definition on the \texttt{.txt} file containing the list of definitions and construct a corresponding function in the \texttt{sketch.py} module that creates a numerical representation of the new definition. Nonetheless, it is a crucial step in increasing the scope of the problems that can be stated for \geosolver{}.

We introduced new definitions with the goal of declaring premises of problems that were not available before but could be managed by the original predicates, of declaring premises of problems that demand the new predicates (de facto including them in the reasoning possibilities of the \geosolver{}), or of making previous definitions more flexible, either reducing conditions for their use or making the statements each definition adds more strict. We introduced the following definitions:

\begin{itemize}
    \item \textbf{on\_pline0.} Similar to the previously existing \texttt{on\_pline}, but drops the requirement that the parallel lines are distinct, allowing one to have overlapping parallel lines.
    \item \textbf{iso\_triangle0.} Similar to the previously existing \texttt{iso\_triangle}, generates the three vertices of an isosceles triangle but adds to the proof state only the fact that the triangle has two equal sides, not also a pair of equal angles as well. This should be proved.
    \item \textbf{iso\_triangle\_vertex.} As in the previous definition, it creates only the apex of an isosceles triangle, but it is weaker than the original definition \texttt{on\_bline}, which included both the congruence of a pair of sides and of a pair of angles into the proof state. This only adds congruence of a pair of sides.
    \item \textbf{iso\_triangle\_vertex\_angle.} This is the complementary definition to \texttt{iso\_triangle\_vertex}, which creates the apex of an isosceles triangle but only adds the statement of the angle congruence to the proof state.
    \item \textbf{on\_aline0.} This new construction of an angle equivalence adds a configuration that did not exist before. Namely, given an intersecting pair of lines, a third line, and a free point, it creates a new point such that the line through this point and the previously existing free point forms an angle with the third line that is congruent to the angle between the other pair of lines. It is more flexible and includes the previous definitions \texttt{on\_aline} and \texttt{angle\_mirror} as special cases.
    \item \textbf{eqratio.} This new construction is the first that allows the prescription of an \texttt{eqratio} predicate, that is, to deal with a premise that two pairs of segments have an equal ratio between them. As for \texttt{on\_aline0}, it takes seven points as arguments (two free existing segments and a free point) to create a new one that completes a fourth segment.
    \item \textbf{eqratio6.} This is another new construction created to insert \texttt{eqratio} conditions that can be problem premises. In this case, it creates a point that will show up twice in the same fraction of the proportion equivalence, which is usually used to {\textit{split}} a given segment with the proportion of a pair of segments. As such, the function takes only six points as arguments.
    \item \textbf{rconst.} This construction was created to insert a \texttt{rconst} predicate as a premise of a problem, that is, a premise that says a pair of segments has a prescribed ratio. It takes one segment and a free point as arguments to construct a fourth point completing the second segment in the ratio.
    \item \textbf{rconst2.} Similarly to \texttt{eqratio6}, this construction allows for an instantiation of \texttt{rconst} in which a point shows up in both segments of the prescribed ratio. It is particularly used for problems which ask for a segment to be split in a ratio of $p:q$.
    \item \textbf{aconst.} This is a new construction for prescribing an angle between two lines with a fixed value without the need to specify the angle vertex. It is the first construction to implement the \texttt{aconst} predicate explicitly.
    \item \textbf{s\_angle.} The original AlphaGeometry had a \texttt{s\_angle} definition, that actually called an \texttt{aconst} predicate on the background. We harmonized it with the overall functioning of the engine. It prescribes an angle of a fixed value with a fixed vertex, and it can be realized by \texttt{aconst} if one uses the intersecting point to describe the pair of lines.
    \item \textbf{lconst.} This construction was created as the entry point for length information into the engine. Given a point, it creates a new point such that both points form a segment with the prescribed length. This is important in order to write problems with information on the length of segments and also to allow a way for length information to enter the problem. In classical geometry terms, it is the only way to introduce the scale, or the unitary segment, into the problem.
\end{itemize}

\subsubsection{Adding New Rules}
\label{subsec:adding_new_rules}

Increasing the number of predicates and definitions is the way to increase the number of geometric problems the engine can reason about. Such new problems can then be constructed and run, but most likely, no solution will be found since the appropriate rules are missing. In order for these problems to be solved, the reasoning side of the engine itself must be improved, and such improvements can take many forms.

We recall that in the original engine, conclusions can be reached by three different paths: application of rules described in the \texttt{rules.txt} file, the running of the algebraic reasoning module, and resolutions that are made on the go through hard-built, not-described functions, that we called intrinsic rules.

The intrinsic rules were created to operate actions humans tend to overlook, such as verifying that if a point lies on a line, it can be an argument for the definition of the line, and that gives rise to a statement of collinearity or that two perpendicular lines form an angle of $90^\circ$ between them, etc. They exist for expediency and for efficiency of execution in some cases, but results from such operations do show up in proofs, as they interfere with the dependency structure, even if they were not mentioned in the original AlphaGeometry paper. Our philosophy towards them was to localize and label them, including mentioning them on the proofs, in order to make their usage explicit, but to rely on them as little as possible in terms of reasoning steps.

The AR module was not explicitly described in the codebase, in the sense that there was no separate list of which facts are turned into equations in the algebraic solver. We altered the algebraic module in the process of inserting new predicates so that they could interact with the algebra operations, see Section \ref{subsec:modifying_ar}.

The \texttt{rules.txt} file is the most explicitly written piece of the reasoning engine. Typically, each line of the file gives you a theorem, structured as a sequence of predicate-enunciated hypotheses, which, if verified for the given proof state, will include the consequence, also predicate-based, into the proof state as well. This set of rules makes up the deductive database (DD) part of the engine. In principle, including a new theorem in the database should be as simple as translating it into an implication described on statements and writing it into a new line of the file. At first, this was not true because some rules could conflict with the traceback or use predicates that were not finalized. This was fixed by our centralization of the predicates.

Our first approach towards adding new rules was to try to add new high-level theorems as rules (\texttt{r43}-\texttt{r48}, see Appendix~\ref{app:rules} for a list of all rules added), with the hope that this would allow the solving of new problems. This looked like a natural extension of the fact that the entire DDAR engine operates not on a purely axiomatic basis but by specifying a collection of theorems that can be used.

However, no new problems from AG-30 were solved thanks to that addition; although we did manage to simplify some of the other existing proofs, that now could be shortened by appealing to the newly added theorems-as-rules. 

A more fruitful approach focusing only on theorems that arguably are \enquote{close} to axioms, was achieved by probing DDAR with small, controlled geometry problems regarding various simple geometric facts to see what facts it knew. In this way, we observed that some basic facts about circles (\texttt{r49} and \texttt{r50} from Appendix~\ref{app:rules}) were missing. Also, we confirmed that there was no connection between the midpoint predicate and the ratio between the corresponding segments being $\frac{1}{2}$, which made us include \textbf{r51}. These small facts did prove fruitful in expanding the capabilities of the solver, including in the benchmark of IMO problems, see Section~\ref{sec:missing_problems}.

Some theorems that one would like to add may not be simple enough to be written with predicates of a simpler form, as the ones inherited from DDAR. In that case, more complex predicates that are potentially less transparent with regard to exactly what they represent have to be designed.

For example, the Pythagorean Theorem is a line in the rules file: 
\begin{center}
    \texttt{PythagorasPremises a b c => PythagorasConclusions a b c}.
\end{center}
Then, when we are checking and performing the resolution of \verb|PythagorasPremise a b c|, the predicate function automatically finds its suitable premises if they are already in the proof state. For its functioning, as we already check numerically every statement before checking it symbolically to feed the cache (see Subsection \ref{subsec:matching}), the solver only needs to check whether the values of the symbols are deducible by the system, but we don't need to really resolve equations to get the values.

\subsubsection{Adding New Equations to AR}
\label{subsec:adding_ar_equations}

\begin{enumerate}
    \item  Following the reorganization of the code, as outlined in Section~\ref{sec:coder_friendliness} on Coder Friendliness, it is easy to list all rules for the addition of equations into AR. Equations are added to the systems of equations at the addition of new statements by calling a \texttt{\_prep\_ar} method. We have the following instances where that happens:
        \begin{itemize}
            \item \texttt{cong A B C D}: adds the equation $\log AB=\log CD$ to the table of ratios.
            \item \texttt{aconst A B C D r}: adds the equation $d(CD)=d(AB)+r$ to the table of angles, where $r$ is any real number that will be taken $\mod \pi$ once the statement is added to the proof state, during the DDARN inference loop.
            \item \texttt{lconst A B l}: adds the equation $\log AB=\log l$ to the table of ratios, where $l$ is a positive number.
            \item \texttt{eqangle A B C D E F G H}: adds the equations $d(CD)-d(AB)=d(GH)-d(EF)$ to the table of angles.
            \item \texttt{eqratio A B C D E F G H}: adds $\log AB-\log CD=\log EF-\log GH$ to the table of ratios.
            \item \texttt{para A B C D}: adds $d(AB)=d(CD)$ to the table of angles.
            \item \texttt{perp A B C D}: adds $d(CD)=d(AB)+\frac{\pi}{2}$ to the table of angles.
        \end{itemize}
    We note that the original codebase contained separate tables of lengths and ratios. We merged them, which resulted in cleaner code, as detailed in Subsection~\ref{subsec:modifying_ar}, and consequently, only a table of ratios is available, referenced above, which includes all lengths. Of this list, the addition of the equations referring to \texttt{cong} statements was moved from the lengths table, used in AlghaGeometry, while the ones referring to \texttt{aconst} and \texttt{lconst} statements are new additions we added to the table of ratios.

\item Further, it is now easy to add new predicates to the engine given the \texttt{Predicate} class we introduced, see Subsection~\ref{subsec:predicates}. This, in turn, allowed us to easily add the predicates that allow the Pythagorean theorem to be used by DDARN.

\end{enumerate}

\section{Coder Friendliness}
\label{sec:coder_friendliness}

A significant amount of our time was dedicated to refactoring the codebase of AlphaGeometry to make the inner workings of its 16,000-line, highly entangled, complex code easier to work with, to eliminate bugs, and to add low-level functionalities that make the user's interaction with AlphaGeometry easier, and on which the previous Section~\ref{subsec:problem_scope} built.

With the exception of the preliminary numerical checking and caching of statements mentioned in Subsection~\ref{subsec:agentic}, this refactoring effort had no direct impact on the reasoning engine. 

Each subsection below sums up the different types of changes and additions we made.

\subsection{Overall Foundational Design}
\label{subsec:foundational_design}

Although we achieved improvements on the reasoning part of the engine, a necessary step to make that possible, and one that took most of the time, was to refactor the original codebase into something more manageable.

The first step, and one that brought \geosolver{} to life, was to separate the AlphaGeometry code into two halves. The first contains everything necessary for the software to solve a problem that does not demand auxiliary constructions, i.e., the processing of a problem as well as the symbolic reasoning structure. The other half is the code necessary for the implementation of the language model used to generate auxiliary constructions. The subsequently created early version of the \geosolver{} codebase could run independently, with the limitation in solvability verified for DDAR in \citep{trinh_solving_2024}, while the LLM's codebase had to call \geosolver{} as a library in order to operate.

This separation of the code not only turns the LLM and the symbolic solver into two codebases that can evolve in parallel but also contains all dependency complications from the original AlphaGeometry
inside the LLM's codebase, so \geosolver{} could be tested and developed from day one. An added benefit of the separate framework between agents and \geosolver{} is that, as dependencies of the agent become optional for the running of the engine, it can be made lighter to install and run if one only wants to use the reasoning engine, without the addition of new points.

After this first change, there were still many iterations of large refactorings of the AlphaGeometry codebase to increase readability and modularity and hence make improvements and debugging easier. The overall principle was to first find the basic ideas effectively used in the functioning of the engine, separate them into smaller modules and classes, each with fewer responsibilities, and centralize important concepts that were scattered throughout the code.

To support the refactoring process, we added several tests. In particular, we have tests for all of the most important rules that are used.

\subsection{Agentic Support}
\label{subsec:agentic}

The reorganization and the modularity we added to the code allow for better communication between \geosolver{} and other software through an interface of what we call \enquote{agents}, as mentioned earlier. As of now, the existing agents are subclasses of the larger \texttt{DeductiveAgent} class inside \geosolver{} itself. 

The currently existing agents are DDARN (which uses breadth-first-search), HumanAgent (that allows human control of the solving process), and flemmard (a dummy agent that does not try to apply anything after building the problem, and is good for testing the parts of the code beyond the reasoning), which we explain below.

We allow the possibility of adding similar classes that allow LLMs, such as the original transformer model from AlphaGeometry, to operate through an agent.

In an agentic setting, important mechanisms are 1) how different agents can interact with each other, 2) what each agent observes and what actions it can take.

We have implemented this by exporting the \enquote{proof state}, carrying all the information about the statements derived by the reasoning at any given moment (agents also accept the file with the rules as input, in case they have to do derivation themselves).

To interact with the proof state, the agent needs to be designed with observation functions, which can directly add new clauses to the problem, match a theorem, add or change a dependency, or check a goal. The functions can be customized to access other information about the proof state, such as the current geometrical graphics of the problem and its premises.

In terms of code, the HumanAgent is an example of an agent that explores all the functionalities of the \texttt{ProofState} class.

\begin{figure}
    \centering
    \includegraphics[width=\linewidth]{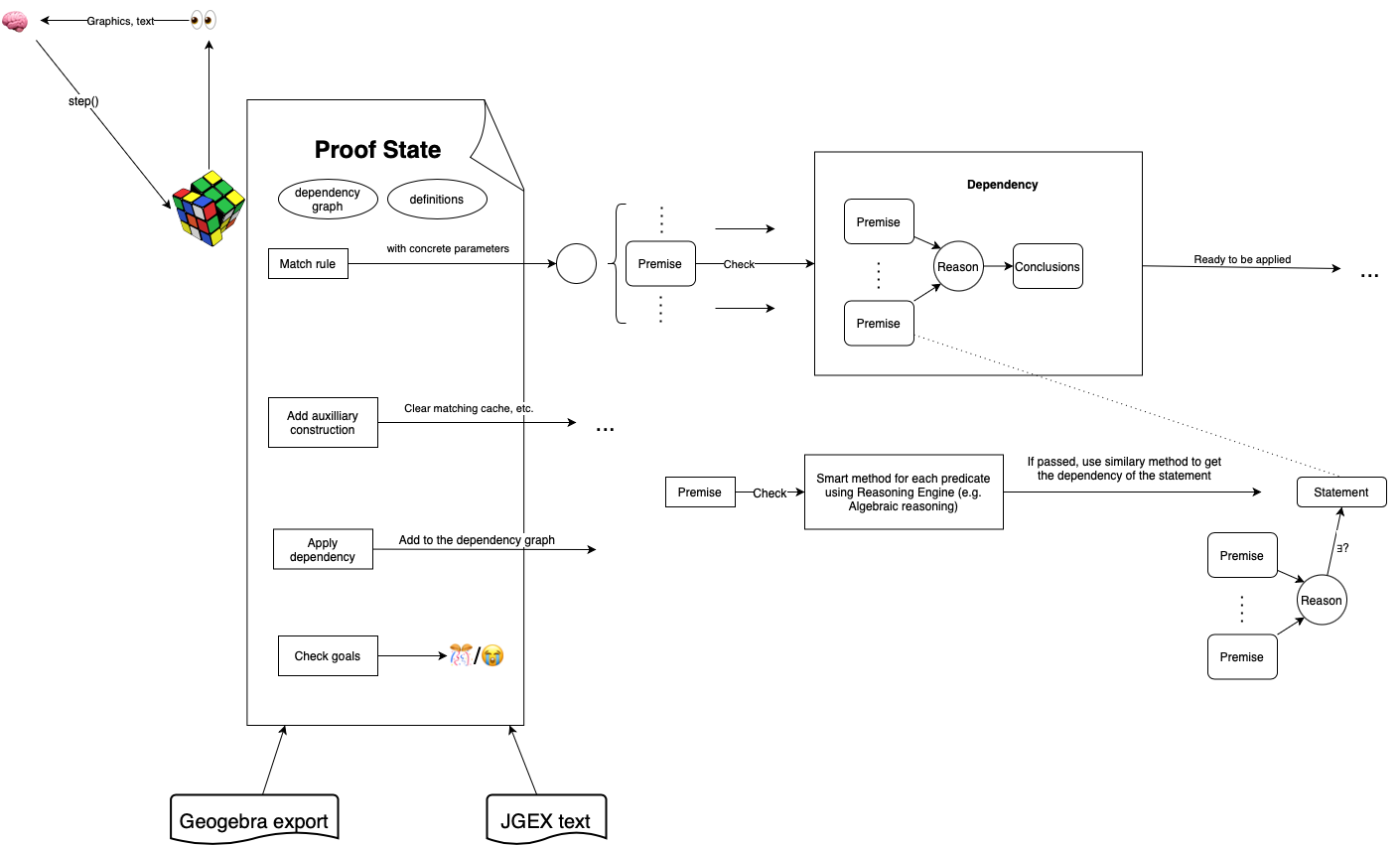}
    \caption{Overview of interactions between different components of \geosolver.}
    \label{fig:overview_inter}
\end{figure}
As shown in Figure \ref{fig:overview_inter}, the interactions pertain to three parts: the agent, the observation functions and the proof states.
A run loop generates steps where, at each step, the agent can manipulate the proof state by matching rules, adding dependencies, and adding clauses that describe auxiliary constructions.

\subsection{Improved Visualization and Debugging Capabilities}
\label{subsec:debugging}

One important contribution to the understanding of the functioning of the code was the visualization of two structures that were implicit and entangled in the original AlphaGeometry code: the symbols graph and the dependency graph. Originally, they were both defined in the single \texttt{Graph} class, whose definition span almost 3000 lines of code, making it very difficult to understand. Also, there were no visualization capabilities built in, so there were no immediate ways to generate these graphs.

\begin{figure}
    \centering
    \includegraphics[width=15cm]{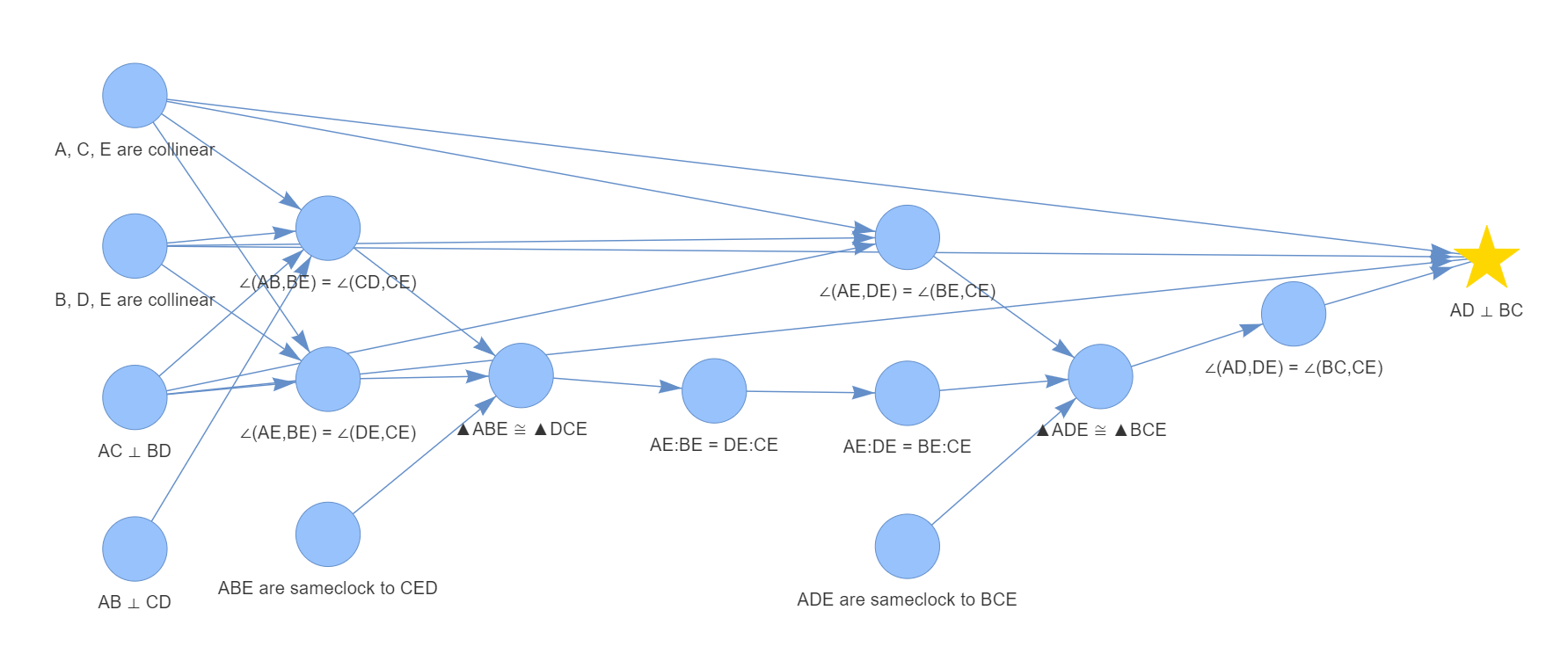}
    \caption{
    \label{fig:dependency_graph}
    A depiction of the reduced dependency graph containing the proof of a simple problem solved by \geosolver (see Appendix~\ref{app:simple_problem} for a formal statement of the problem). This reduced graph only shows when a problem can be solved, which allows us to prune the full dependency graph of any nodes that did not contribute to the solution of the problem.}
\end{figure}

As the name suggests, the dependency graph depicts the dependency structure of the proof, namely what premises were derived, on what other premises they rely on, and what the mechanism that allowed for that derivation is. It is an and-or graph, with nodes representing statements and directed edges representing {\textit{dependencies}} between the statements. Each dependency can consist of potentially multiple, or even zero, incoming statements and one outgoing statement, in addition to a reason citing the rule allowing the deduction of the outgoing statement from the incoming ones. There are two depictions of the dependency graph: One can see the \emph{full dependency graph}, which shows all the dependencies between all the statements that DDARN derives (which appears if a solution is not found at all), or one sees the \emph{reduced dependency graph}, available only in case where a problem was solved, where we traverse the full dependency graph and exclude any nodes (and corresponding edges) that are not relevant to the found solution, using the traceback.

\begin{figure}
    \centering
    \includegraphics[width=14cm]{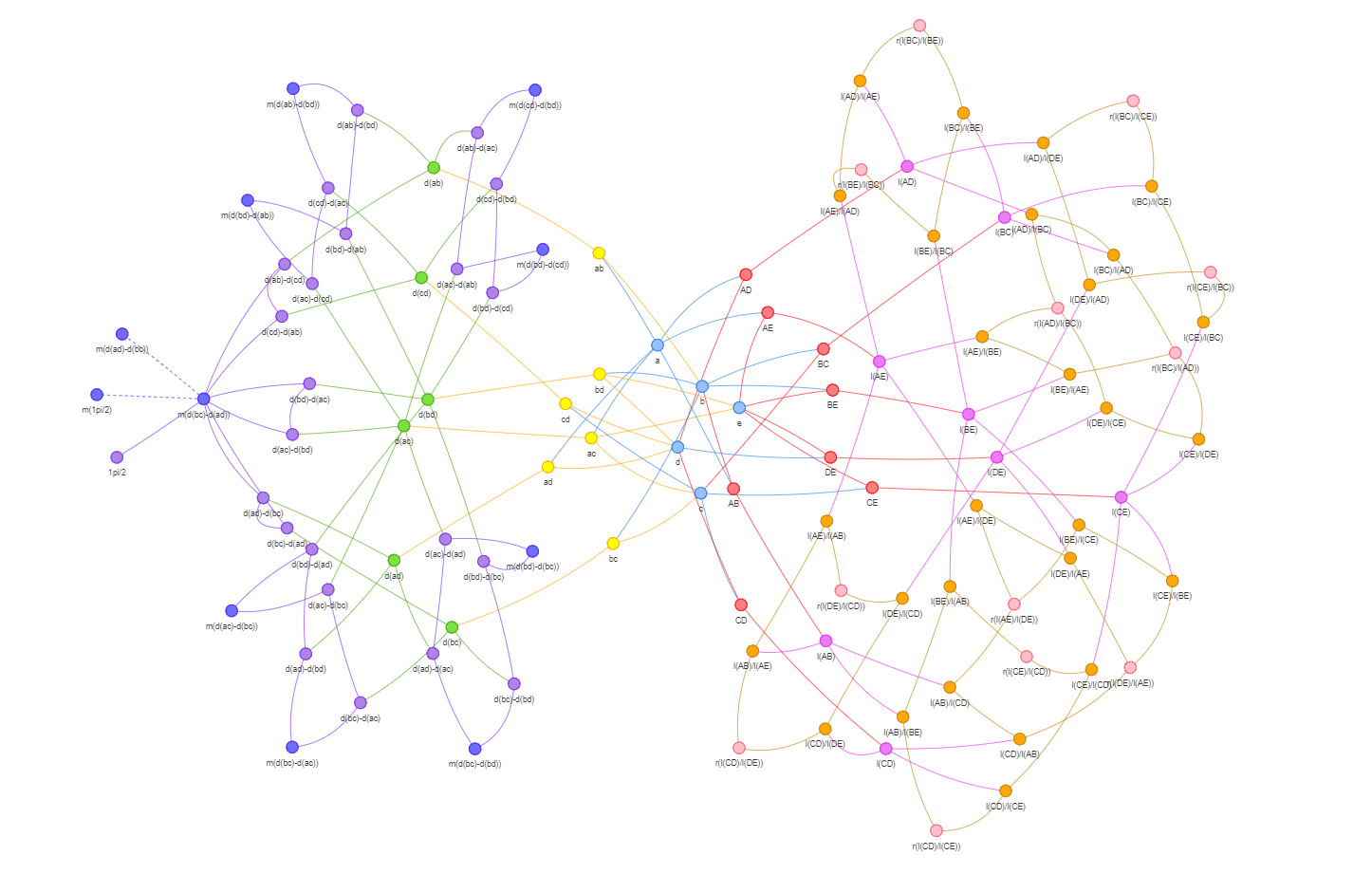}
    \caption{
    \label{fig:complex_symbols_graph}
    A depiction of what the symbols graph of a small problem that could be solved by DDAR in the original AlphaGeometry, with two components, one for angles and one for ratios, looks like (see Appendix~\ref{app:simple_problem} for a formal statement of the problem). Obtaining this figure already relies on code changes to AlphaGeometry to extract the symbols graph while not exhibiting any of the further refactorings we made to the storing of geometric objects.}
\end{figure}

Less obvious is the symbols graph, which depicts all the geometric objects created in the construction of the problem and the development of the proof, and if they are composed of smaller objects, they will be connected to those basic objects. 

\subsubsection{Dependency Graph}
\label{subsec:dependency_structure}

The biggest feature that was missing that we encountered in the AlphaGeometry code, was clear documentation of the dependency structure: a datastructure that collects the statements and their dependencies in a proof. Although the original paper by~\citet{trinh_solving_2024} mentions the need for a traceback to simplify proofs and to weed out statements that were derived during the DDAR loop but that are not necessary for the proof, it does not touch on the complexity of the functions located throughout the code dedicated to registering and sorting out the dependencies of statements generated in the process.

Such dependencies are now stored in what is called the \texttt{Dependency} class, and the collection of all dependencies forms the (full) dependency graph. A dependency is typically built when a statement is checked symbolically, a rule is applied, or exceptionally when the symbols graph is synthesised by merging lines or circles, which happens when the engine finds out three points (which would correspond to three lines in the symbols graph), are actually collinear. It consists of a justification (such as the name of the rule), its premises and a conclusion. Also, when a dependency is added to the dependency graph, a function related to the conclusion predicate can be triggered, and the inner states of the AR module and the symbols graph can be modified.

\begin{figure}
    \centering
    \includegraphics[scale=0.42]{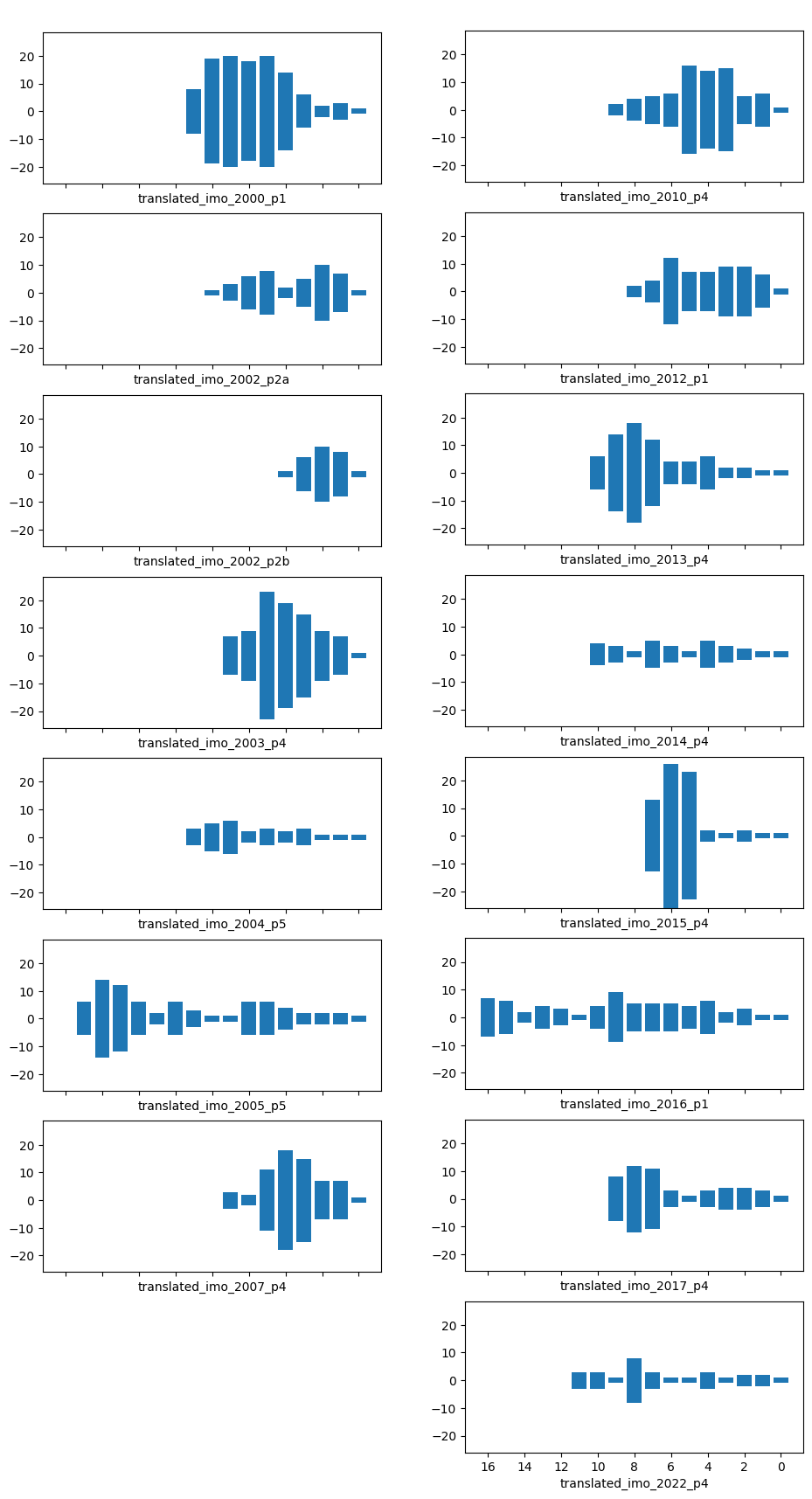}
    \caption{The width at every depth of the dependency graphs, with levels ordered from statements that are closer to the goal in the proof, for all the 15 problems in the \texttt{imo\_ag\_30} benchmark that can be solved by \geosolver's DDARN module (with breadth-first-search).}
\end{figure}

\subsubsection{Symbols Graph}

Initially, the symbols graph contained many kinds of objects: points, segments, directions, angles, ratios, measurements of angles, values of ratios, the numerical value of those measures, and circles. This myriad of classes of geometric objects was necessary because many implicit derivations of facts took place through identifications of nodes on the symbols graph. In our work, we moved a big part of the reasoning responsibilities of the symbols graph, namely the ones involving congruences of measures, to the AR module, and some others were given to explicit rules. For DDARN, the cleaned-up symbols graph only stores points, lines, and circles, and the only reasoning it carries concerns detecting and storing when new points are found in already existing lines and circles (i.e. collinearity and concyclicity). It should be stressed that such identification is only accepted by the engine if there is a symbolic justification for it, the numeric recognition that a point lies on a given line on circle is not enough for the registering of that information in the symbols graph.

\begin{figure}
    \centering
    \includegraphics[height=7cm]{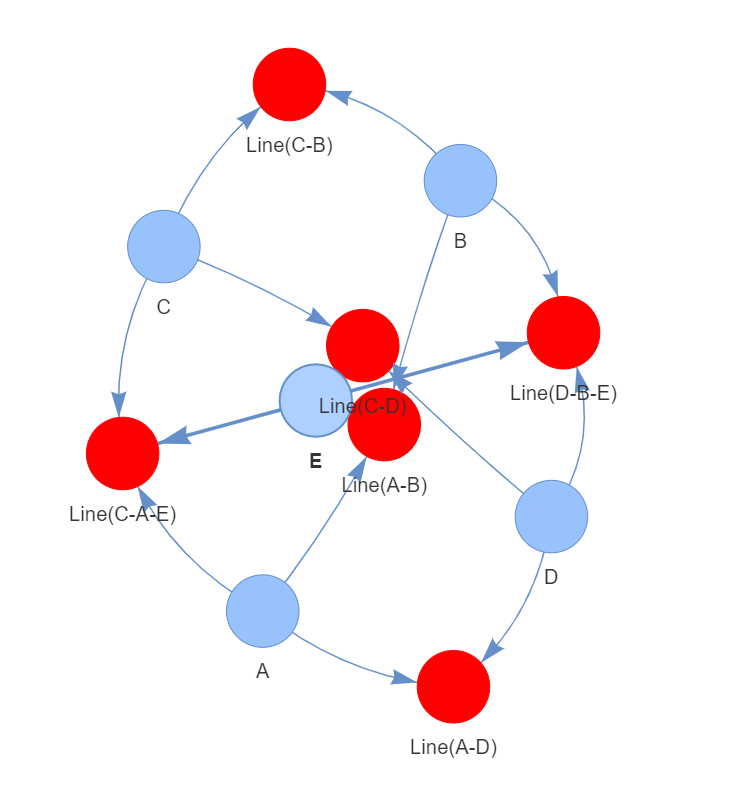}
    \caption{The DDARN symbols graph of the same problem as in Figure~\ref{fig:complex_symbols_graph}, generated by \geosolver. The essential relationships present in the problem can be easily read from the symbols graph.}
\end{figure}

\subsubsection{Using the Graphs}

Visualizing this information in these graphs and inspecting them had important consequences. It turned out that they are one of the most comprehensive visual depictions one can have of the reasoning of the engine, to the point that we found out that the original \texttt{s\_angle} definition was processed through a hidden, not entirely functional, \texttt{aconst} predicate because that dependency surprisingly showed up in dependency graphs (but not in the proof).

Both the symbols graph and the dependency graph were crucial concepts to understand the underlying structure of the reasoning beyond the information contained in generated proofs. In addition, having these visualization tools that contain a lot more information on the reasoning of the engine than the proofs was crucial for debugging, as it pointed us to incorrect or superfluous connections that were made by the engine. This information, in turn, made it possible to make various additions to fix these problems.

\subsection{Reasoning Engine Improvements}
\label{subsec:engine_improve}

The guiding principle of DDAR was to be able to apply theorems to a problem in a breadth-first search (BFS). To execute this idea, a series of technical challenges have arisen regarding how to structure and store the mathematical information and how to implement the search. A big part of our effort focused on creating good, flexible code structures that could execute the breadth-first strategy that DDARN employs in a better way than AlphaGeometry. In order to do so, we implemented and modified the structures described below.

\subsubsection{Predicates}
\label{subsec:predicates}

One major step in giving the code flexibility was to create the \texttt{Predicate} class and to organize the most fundamental reasoning terminology of \geosolver{} as a collection of subclasses, with a structure that can be expanded if needed and easily modified. Predicates are more complex than definitions, so a \texttt{Predicate} class is more complex than a definition statement on the \texttt{defs.txt} file, but the philosophy we followed was that the former should be as easily found and modified as the latter.

A \texttt{Predicate} class contains methods to:
\begin{itemize}
    \item Put the arguments of the predicate in a canonical order (\texttt{preparse}).
    \item Parse its arguments from strings to geometrical points (\texttt{parse}).
    \item Find a dependency structure that can justify the existence of that predicate and do the symbolic check (\texttt{why}, \texttt{add}).
    \item Check numerically for the validity of that predicate, do the numerical check (\texttt{check\textunderscore numerical}).
    \item Process the representation of that predicate when writing the proof, generating diagrams, or receiving and passing information to the LLM (\texttt{pretty}).
\end{itemize}

This list is typical but not exhaustive, as some predicates may demand extra processing functions in the background. 

For example, in order to apply Pythagoras' theorem, as can be seen in the \texttt{rules.txt} file pertaining to this example, two predicates were created, \texttt{PythagoreanPremises} to represent the premises (the presence of a right-angled triangle), and \texttt{PythagoreanConclusions} to represent the conclusion (the sum of the squares of the legs). The \texttt{PythagorasConclusions} predicate has a method to extract the distance of the missing side of the triangle (if there are two known sides) numerically and add the corresponding \texttt{lconst} statement to the proof state, avoiding the need for manipulating an equation. This method, of course, does not have an equivalent in other predicates, but the fact that predicates are classes adds this flexibility to their definition while keeping all clauses needed in one place.

\subsubsection{Algebra Reasoning}
\label{subsec:modifying_ar}

One of the main contributions of the authors of~\citep{trinh_solving_2024} to the development of a geometry reasoning engine was the addition of the algebraic reasoning module (AR) in tandem with the deductive database (DD) resulting in the DDAR solver, that increases the range of statements found compared to DD. AR is a symbolic engine for checking and getting the justification of statements based on an internal set of linear equations.

The original engine was built on three sets of linear equations, called tables: one storing information about angles, one storing information about ratios, and one storing information about lengths. Each {\textit{internal linear equation}} is of the form $y=\sum q_i x_i+\text{constant}$, where only the $q_i$'s are stored explicitly. When a new equation is given, it is simplified so that it contains only free variables on the right-hand side, with one variable, $y$, to depend on others, if necessary.

The goal of the AR module is to find new equations of the form $\sum b_ix_i=\text{Constant}$, corresponding to new predicates not present in the previous equations. This involves resolving the dependency structure of the predicate found to minimize proofs at the time of the traceback. Different procedures are applied at each of those stages.

To find the new statements, as described in the original work by~\citep{trinh_solving_2024}, the engine applies Gaussian elimination at each table, a process that can result in a new statement or not. Each table stores information about all occurrences of predicates feeding it, but not all of them necessarily relate to the new statement found. Once a goal is found, to weed out unrelated statements in the table, the engine uses linear programming techniques to find the minimal system that satisfies the new statement, aiming at a shorter proof.

All those procedures are part of AlphaGeometry. In order to have a clearer algebraic module that is easier to understand and modify and that is more robust against mistakes, we made minor changes.

The most notable one was the removal of the length table. The lengths of segments were arguments in both the table of lengths and the one of ratios (in this case, ratios are linearized by the application of logarithms as described in~\citep{trinh_solving_2024}). The table of lengths, though, had only simple information about the equality of segment lengths, information that we easily incorporated into the table of ratios without further negative consequences, reducing the number of tables to two.

As this table now has to deal with information on constant lengths, but the entries are logarithms of lengths for the linearization of ratios, we end up having to deal with the logarithms of constants. To minimize numerical instabilities and simplify symbolic manipulations, instead of storing those throughout the Gaussian elimination process, we retrieve them through numerical checks whenever the proof state has dependencies supporting the corresponding \texttt{lconst} statements(see Subsection~\ref{subsec:adding_ar_equations}). 

\subsubsection{Matching}
\label{subsec:matching}

At first look, a BFS search for an automatic theorem prover may look as costly as the number of theorems it has to go through while performing the search. Our experience with AlphaGeometry shows that this is only part of the story. In fact, the process of using theorems can be separated into one that matches the theorem, that is, looks for arguments to fit the predicates in the hypothesis of the theorem, and one that applies the theorem generating a new statement from the theorem's conclusion. The matching is the combinatorially expensive step of the process, and it can make some theorems consume significantly more resources than others.

Generally, to reduce the amount of resources needed for running a loop in a symbolic solver that is responsible for the application of theorems, one could either improve the search strategy, which has been done, for example, by the use of reinforcement learning and Monte Carlo tree-search in unrelated earlier work, e.g. by~\citet{lample2022hypertree}, but this is a more difficult task. A different strategy that is more easily implemented is to reduce the space of possible mappings for the theorems.

We implemented the latter strategy in two ways. First, we implemented a {\textit{matching cache}}. The idea here is that the matching only takes into account real statements, instead of all combinations of arguments present in the problem, by numerically checking all possibly matched statements and caching the true ones in the disk at the beginning of the problem. This way, the full cost of matching is concentrated at the beginning of the process of solving the problem for the first time, which makes the time spent in the BFS loop itself really short. This reduces the overall time to solve a problem by a bit, and makes debugging and improving the code a lot faster, even if one tests features on large theorems, as the matching time for building the cache is only spent in the first iteration of the problem.

Another strategy we used for reducing the time spent in matching was to reorder the statements in the hypothesis of theorems in the \texttt{rules.txt} file in such a way that the first statements are the ones with fewer arguments. 
The guiding principle was: \emph{if} the rule checking is going to fail, we want it to fail faster, i.e. without checking all the premises. We thus put the premises with fewer arguments ahead in the checking loop, as these are resolved faster. Thus, time doesn't need to be spent on the remaining premises with more arguments, which take longer to resolve.

This way, failures of theorems happen faster, saving unnecessary time spent on slower matching statements with more arguments. This also has the advantage of being easily tested, as one can perform experiments simply by generating an alternative \texttt{rules.txt} file. 
Testing a reorganization of the rules' hypothesis against the \texttt{imo\_ag\_30} benchmark shows a reduction of a little more than 10\% on those problems, see Figure~\ref{fig:time_spent}.

\begin{figure}
\centering
\begin{subfigure}{\textwidth}
    \centering
    \includegraphics[height=7cm]{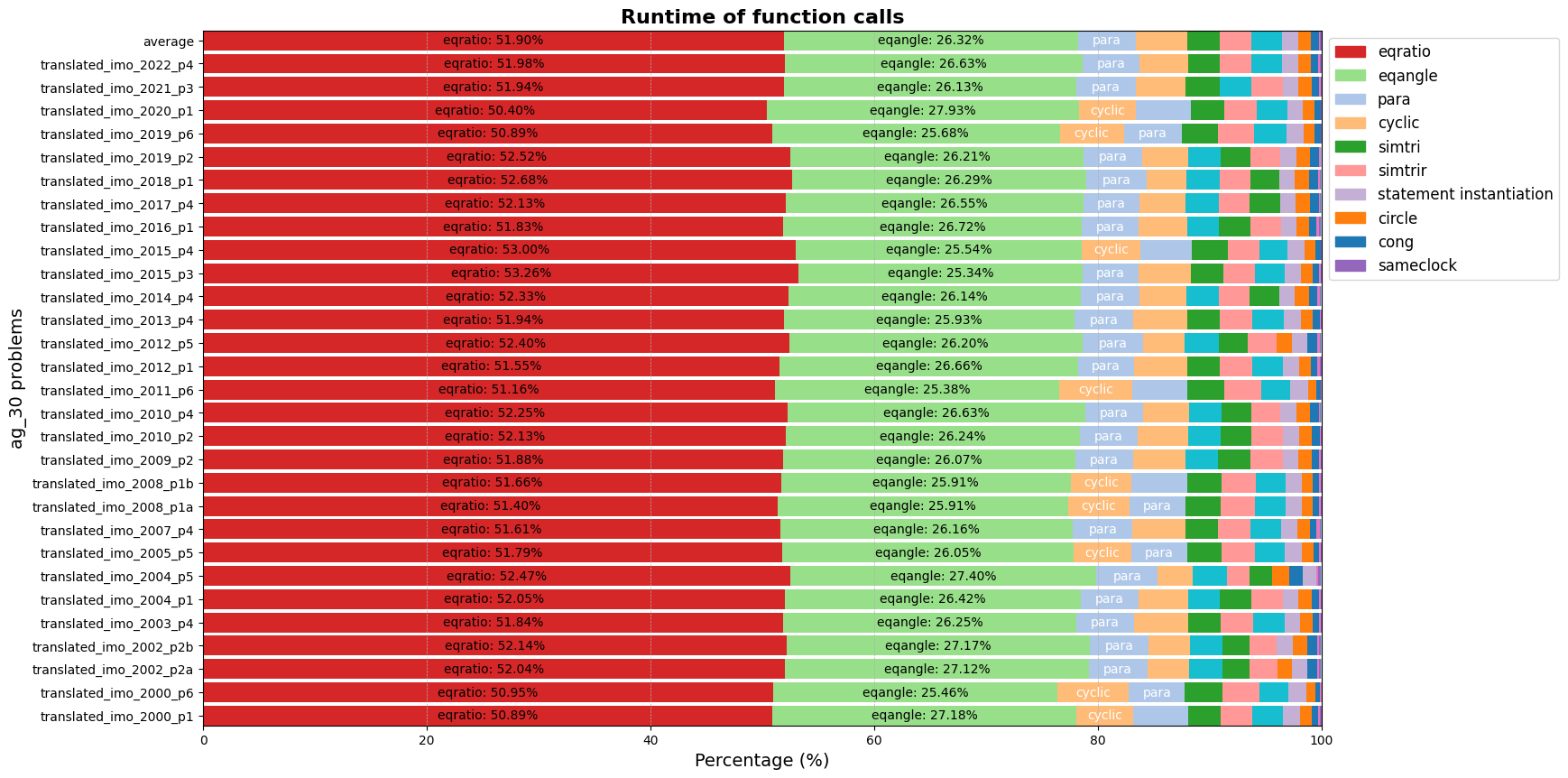}
    \caption{Percentage distribution of time spent on matching predicates with rules written in the old order, with a clear prevalence of time spent on the costly \texttt{eqratio} and \texttt{eqangle} predicates.}
\end{subfigure}

\vspace{1em} %

\begin{subfigure}{\textwidth}
    \centering
    \includegraphics[height=7cm]{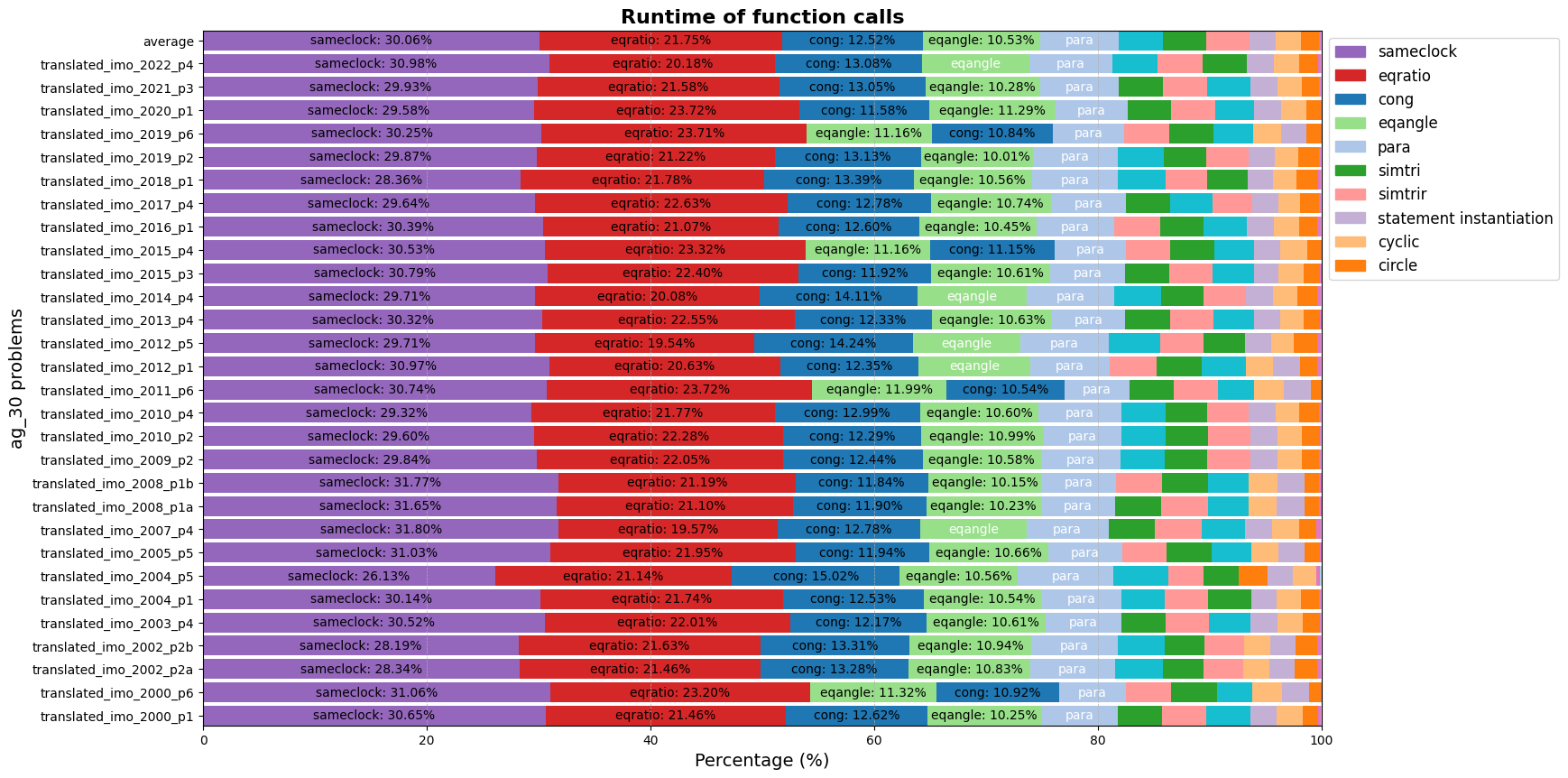}
    \caption{Percentage distribution of time spent on matching predicates with rules written in the new order, with more time spent on the quicker \texttt{same\_clock} predicate.}
\end{subfigure}

\caption{
\label{fig:time_spent}
The profiling of the level of functions where the matching of predicates occurs in the code, for the problems in the \texttt{imo\_ag\_30} benchmark. It clearly shows that most of the time is spent on matching specific predicates, which do not depend on the nature of the problem. More time spent on quicker predicates, as can be seen at the bottom, results in a shorter overall time for solving problems.}
\end{figure}

\section{Reproducibility}
\label{sec:reproducibility}

For a third-party checking of the experimental results presented in~\citep{trinh_solving_2024}, given the nature of the software, there are two different aspects to be considered: on one hand, one should be able to run the problems on AlphaGeometry's code in order to get the same outcomes, on the other hand, someone with the proper mathematical knowledge should be able to know which steps were taken by the engine and verify if the proof written was mathematically correct. We identified issues in AlphaGeometry (and DDAR specifically, since the LLM is secondary to these issues) regarding both aspects and tried to introduce improvements in both the possibility of reproducing results and verifying their soundness. We elaborate on these below and highlight how the new additions made in previous sections aid reproducibility.

When it comes to understanding and verifying proofs, DDAR presented some issues. The most visible one was that some proof steps were presented without justification; that is, the recorded proofs did not explicitly state which rules justified each implication. While it is true that it is common in mathematics that arguments are sometimes presented with little to no justification, assuming the reader can complete the argument with previous knowledge, an automatic prover has to satisfy a higher level of rigour because it makes it hard to understand the inner workings of a symbolic engine (and therefore to anticipate potential weaknesses) from proof trace inspection if there are gaps. In the case of DDAR, this issue is even deeper, as some steps in the proof use intrinsic rules, which are not made explicit anywhere.

One of our first tasks was to try to fill this gap (which led us to uncover the dependency structure of proofs, mentioned in Subsection~\ref{subsec:dependency_structure}). We were only able to have fully descriptive proofs once we rebuilt the dependency graph of the problems, with full control of how dependencies, in particular the ones generated in the background, were transmitted to statements. By contrast, in DDARN a user can fully evaluate the statements derived at each step and decide if any given step is sound or not.

A more subtle issue we found when trying to run problems with DDAR at different times was that the proofs written are not deterministic; that is, proofs change between iterations even without changes in the parameters. (We emphasize that this happens in DDAR alone, even where proofs are carried out without the intervention of the LLM, which naturally introduces nondeterminism.) Investigating this unexpected phenomenon, we found that the assignment of coordinates to points during the building of the problem contained randomness, and that seemed to be the only source of randomness in the DDAR pipeline. In principle, this randomness should not change the symbolic nature of the proof state. However, in practice, it can alter the predicates that are obtained from numerical checks of the diagram built at the instantiation of the problem. In particular, it can generally, with a $50\%$ chance, change the orientation of triangles.

Now, there was no explicit condition on the orientation of vertices of triangles to be found on the rules or to be identified as a predicate. However, when checking conditions for the similarity and congruence of triangles, because of the use of the concept of total angles in DDAR, the orientation of vertices of triangles is relevant. Indeed, hidden inside the specific theorem matching functions, one could find checks for the orientation of vertices of the triangles being compared (see the \texttt{same\_clock} functions, which turn these hidden functions in explicit predicates, mentioned in Subsection~\ref{subsec:adding_new_predicates}), and those influenced the final shape of the proofs.

In more extreme cases, for example, in the proof of Problem 2 of the IMO 2009 exam, given in the Supplementary Material to the AlphaGeometry paper~\cite{trinh_solving_2024}, page 29-30, the very correctness of the proof depends on a choice made at random by the software. Specifically, one of the auxiliary points described there, $D$, is given by the intersection of two circles. Now, a pair of circles intersect at two points, and in that case, the solver makes the choice of which point to use at random (without making it explicit). It so happens that one can find errors in the written proof if the point not chosen (but that also satisfies the defining conditions of $D$) were to be considered by the AlphaGeometry software: In that case, the statement proved in step 11 of the proof ($\angle MLK = \angle DML$) is false if one considers full angles, and the statement proved in step 15 ($\angle (LD, CO)=\angle QBO$) is false under any formal definition of angles. Thus, the proof in the Supplementary Material, as written, must have been arrived at from the choice of the correct point in the intersection of the circles, although this fact is not explicitly noted.

To address those issues, we took extra care to make every step explicit. First, we added the mentioned \texttt{same\_clock} predicate as a numerically checked predicate and enunciated it as a hypothesis on the rules concerning the similarity and congruence of triangles. As a consequence, the proof will explicitly show the choice made and the use of facts about the orientation of the vertices of the triangle in the proof. 

Indeed, combining that with the functionality we added for checking multiple goals, one can steer the random choices of the orientation of triangles in the problem, and doing that for Problem 2 of the 2009 IMO exam, one can verify that DDAR would not have been able to find a solution for the problem with the other different choice of $D$, even if both provide the same statements to the proof state by the prescribed auxiliary point.

Separately, to control the randomness of the engine, we introduced a seed that can control the random choices of \geosolver{}, which must be done carefully. The randomization of the prescription of coordinates to points is not a trivial aspect of the building of a problem.
In fact, every time coordinates are prescribed in the construction of a numerical representation of the problem, the engine will check the correctness of the goals, and if a goal is not numerically true, the software will scrap the coordinates and try to assign new ones. This can be repeated a finite number of times before \geosolver{} gives up (this is currently set to be repeated 10000 times, but it only matters that it is a sufficiently large number), in contrast to the original version of the software that used a \texttt{while} loop, risking a potentially infinite loop for problems that had false goals. This is important not only as a sanity check but also because some assumptions needed for a problem are not contemplated by the language available to \geosolver, for example, an open constraint on the relative sides of two segments or the fact that a point is internal or external to a triangle. If the space of points satisfying those conditions has a positive measure in the space of all possible configurations within the random choices of the problem, the good configuration can be reached randomly, given sufficiently many tries, and that is what the randomization allows. Of course, for testing, it is useful to be able to fix a good seed and have better control of the problem.

\section{The 5 Missing Problems From \texttt{imo\_ag\_30.txt}}
\label{sec:missing_problems}

In AlphaGeometry paper by~\citet{trinh_solving_2024}, the authors compile a collection of 30 problems in the AlphaGeometry format from 27 unique problems from the IMO exams between 2000 and 2022 (the last officially released exam before the writing of the paper). 
The three remaining problems arise from splitting three of the problems in the set of 27 problems into two problems.
From those, the (original) DDAR engine solved 14/30 problems, while the full AlphaGeometry (i.e., DDAR and the LLM working in tandem) at full computing capacity could solve 25/30 problems. The adaptations and solutions to those 30 problems are collected in the Supplementary Material of~\citep{trinh_solving_2024}, their formulation into the language used by AlphaGeometry is contained in the file \texttt{imo\_ag\_30.txt} of the original AlphaGeometry codebase on GitHub.

Studying the five problems that AlphaGeometry was not able to solve (2008 P1B, 2008 P6, 2011 P6, 2019 P2, and 2021 P3) is instructive and provides insights regarding the capabilities and limitations of AlphaGeometry both as a concrete software, with its technical and specific design limitations, and as an overall project of a coordinate agnostic automatic reasoning system. 

Below, we discuss each of these five problems, highlight what was missing for the engine to solve them, or the reason for their non-solvability when that is the case.

One should keep in mind that there is a level of stochasticity in the analysis of why problems fail, as one can only speculate on the precise reasons a probability-based machine such as an LLM takes a specific path in comparison to another one. Also, the axiomatic reasoning developed by DDAR, which is also incorporated in DDARN, despite being deterministic in principle, is highly sensitive to changes in the formulation of problems and rules, so it is to be expected that other modifications different from the ones we proposed can have the same positive outcome as a result.

Problem 2 from IMO 2019 is already discussed in the Extended Data, Figure 4, of the original paper, so it is briefly mentioned here for completeness only.

\subsection{IMO 2008 P1B}
\label{subsec:imo_2008_p1b}

\begin{problem}[Original Formulation of P1 2008 (Evan Chen's Solution Notes)]

Let $H$ be the orthocenter of the acute-angled triangle $ABC$. The circle $\Gamma_A$ centered at the midpoint of $\overline{BC}$ and passing through $H$ intersects the sideline $BC$ at points $A_1$ and $A_2$. Similarly, define the points $B_1$, $B_2$, $C_1$, and $C_2$. Prove that six points $A_1$, $A_2$, $B_1$, $B_2$, $C_1$, $C_2$ are concyclic.
\end{problem}

\paragraph{Explanation of our translation of P1 2008 (with auxiliary points in red):} Consider the triangle $ABC$, and that also
\begin{itemize}
    \item $H$ is the orthocenter of $ABC$.
    \item $D$ is the midpoint of $BC$.
    \item $E$ is the midpoint of $AC$.
    \item $F$ is the midpoint of $AB$.
    \item $A_1$ is the first intersection of the circle of center $D$ through $H$ and line $BC$.
    \item $A_2$ is the second intersection of the circle of center $D$ through $H$ and line $BC$.
    \item $B_1$ is the first intersection of the circle of center $E$ through $H$ and line $AC$.
    \item $B_2$ is the second intersection of the circle of center $E$ through $H$ and line $AC$.
    \item $C_1$ is the first intersection of the circle of center $F$ through $H$ and line $AB$.
    \item $C_2$ is the second intersection of the circle of center $F$ through $H$ and line $AB$.
    \item $\textcolor{red}{O_1}$ is the point on the line perpendicular to $EF$, through $H$, such that the distance from $E$ to $O_1$ is the distance from $E$ to $H$ (the reflection of $H$ across the line $EF$).
    \item $\textcolor{red}{O_2}$ is the point on the line perpendicular to $ED$, through $H$, such that the distance from $E$ to $O_2$ is the distance from $E$ to $H$ (the reflection of $H$ across the line $ED$).
    \item $\textcolor{red}{O_3}$ is the point on the line perpendicular to $FD$, through $H$, such that the distance from $F$ to $O_3$ is the distance from $F$ to $H$ (the reflection of $H$ across the line $FD$).
    \item $\textcolor{red}{O}$ is the center of the circle through $C_1$, $C_2$, and $B_1$.
\end{itemize}
Then, prove that $A_1$, $A_2$, $B_1$, $B_2$, $C_1$, $C_2$ are concyclic.

\paragraph{Formal translation of our problem:} \

\begin{lstlisting}[language=jgex]
a b c = triangle a b c; 
h = orthocenter h a b c; 
d = midpoint d b c; 
e = midpoint e a c;
f = midpoint f a b;
a1 = on_circle a1 d h, on_line a1 b c; 
a2 = on_circle a2 d h, on_line a2 b c; 
b1 = on_circle b1 e h, on_line b1 c a; 
b2 = on_circle b2 e h, on_line b2 c a; 
c1 = on_circle c1 f h, on_line c1 a b; 
c2 = on_circle c2 f h, on_line c2 a b; 
o1 = eqdistance o1 e e h, on_tline o1 h e f; 
o2 = eqdistance o2 e e h, on_tline o2 h e d; 
o3 = eqdistance o3 f f h, on_tline o3 h f d; 
o = circle o c1 c2 b1 ? cyclic c1 c2 b1 b2 a1 a2
\end{lstlisting}

\begin{figure}
    \centering
    \includegraphics[height=10cm,width=12cm]{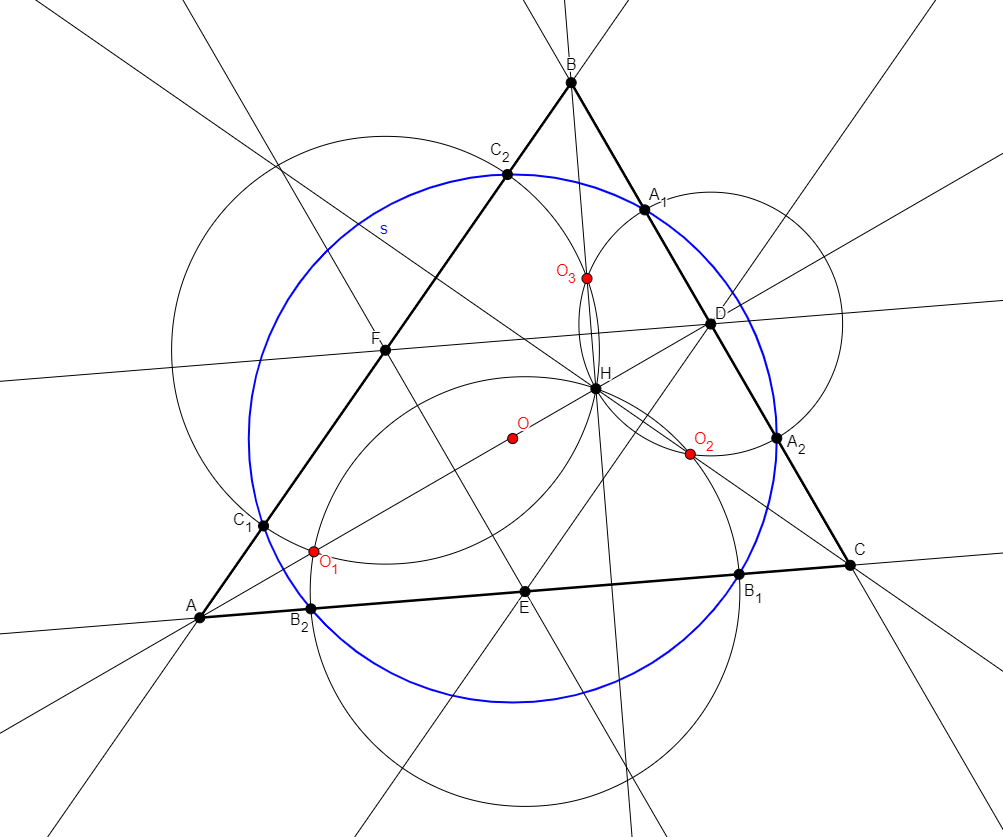}
    \caption{Diagram for the statement of Problem 1 of the IMO 2008 exam.}
\end{figure}

\paragraph{Discussion:} Problem 1 of the 2008 IMO exam asked to prove that a collection of six points ($A_1$, $A_2$, $B_1$, $B_2$, $C_1$, and $C_2$ in the original formulation) lied in the same circle. 
In terms of the internal language of AlphaGeometry, that should be expressed by a \texttt{cyclic} predicate.

In the original formulation by~\citet{trinh_solving_2024}, problem 1 from the 2008 IMO exam was split into two problems, one asking for \texttt{cyclic B1 B2 C1 C2} (2008 P1A) and the other asking for \texttt{cyclic C1 C2 B1 A1} (2008 P1B), choice that was probably motivated by the fact that the \enquote{natural} presentation of the \texttt{cyclic} predicate has four arguments, as any triple of points in general position determines a circle. In case both problems were solved, still some human vision would be needed to understand that proving both facts is enough for the complete problem, as they cover the two possible configurations of quadruples of points distributed pairwise on the sides of a triangle. So the full problem could be proven \enquote{analogously}. Of course, a larger issue is that AlphaGeometry was only able to solve P1A, but not P1B.

Trying to understand what AlphaGeometry missed for this problem, the first thing to observe is that \texttt{cyclic} is a predicate that accepts any list of points, so one can straight away ask for \texttt{cyclic A1 A2 B1 B2 C1 C2} as a goal in the formulation, without the need to split the problem in two. Of course, as AlphaGeometry could not solve P1B, it cannot solve the unified formulation either, even if provided via a new clause with the additional clauses that allowed the solution of P1A.

But observing the clause suggested for the proof of P1A, we can get a good outline of how to proceed in solving this problem: P1A was solved by adding the point $O$ ($\textcolor{red}{O_1}$ in our formulation), the reflection of the orthocenter $H$ of the triangle by a line connecting midpoints $E$ and $F$ of sides $AC$ and $AB$, respectively. Those are exactly the sides where points $B1$, $B2$, $C1$, and $C2$, mentioned in the goal of P1A, lie. One way to prove that all the points mentioned lie in a single cycle would be to reprove P1A for the three pairs of adjacent sides of the triangle and then show that all the circles found are the same.

With this in mind, we prescribed $O1$, $O2$, and $O3$, the reflections of $H$ along the three lines connecting midpoints of sides of the triangle, extending the suggestion given for P1A. This proved not to be enough for DDARN to prove all six points lie on the {\textit{same}} circle, so we also added $O$, the center of the circle through $C1$, $C2$, and $B1$ (and the common center to all circles already detected by the solver analogously to P1A). Then DDARN could find a solution to the problem.

The existence of this solution suggests a better understanding both of the behaviour of the LLM and of the limitations of the original rules of DDAR. For the LLM, it is plausible to assume that, as it did find a good clause $O$ for 2008 P1A in the original run described in~\citep{trinh_solving_2024}, it could propose $O1$, $O2$, and $O3$ in our construction, as they all play the same role, up to a cyclic permutation of the sides of the triangle. The problem, of course, is that there is no evidence that the LLM can profit from such symmetry of points, so it would need to repeat whatever reasoning brought it $O$ thrice, demanding triple resources. This does not mean it could decide to suggest {\emph{our}} $O$, the center of the circle asked in the goal, although that is also a natural point to suggest. Simply making all those decisions would demand too much depth in the search.

On the other hand, we did try to run our new formulation of 2008 P1 with the original set of rules from DDAR, and it still exhausted the DDAR search without finding a proof. Reviewing our proof, on the other hand, reveals the use of rules \texttt{r49} and \texttt{r50} (the need for rule \texttt{r50}, in fact, was detected by studying this very problem); see Appendix~\ref{app:rules}. Those two rules contain very simple facts about circles: that points lying on the same circle are equidistant from the center and that the intersection of the perpendicular bisectors of non-parallel chords of a circle happens at its center. Nonetheless, those facts were unknown to the original DDAR formulation, and their addition is necessary for the solution of some problems involving circumferences and cyclic points, such as 2008 P1.

A full output of the solution of \geosolver{} for this problem can be found in Appendix~\ref{app:full_solution}.

\subsection{IMO 2008 P6}

\begin{problem}[Original Formulation of P6 2008 (Evan Chen's Solution Notes)]

Let $ABCD$ be a convex quadrilateral with $BA\neq BC$. Denote the incircles of triangles $ABC$ and $ADC$ by $\omega_1$ and $\omega_2$ respectively. Suppose that there exists a circle $\omega$ tangent to ray $BA$ beyond $A$ and to the ray $BC$ beyond $C$, which is also tangent to the lines $AD$ and $CD$. Prove that the common external tangents to $\omega_1$ and $\omega_2$ intersect on $\omega$.
\end{problem}

\paragraph{Explanation of AlphaGeometry's translation of P6 2008:} Consider the triangle $XYZ$ with prescribed coordinates (they will determine the circle $\omega$), and also that
\begin{itemize}
    \item $O$ is the center of the circle through $X$, $Y$, and $Z$.
    \item $W$ is a point on the circle with center $O$ through $X$, with prescribed coordinates.
    \item $A$ is the intersection of the line perpendicular to $OZ$ at $Z$ with the line perpendicular to $OX$ at $X$ (so $AZ$ is tangent to $\omega$ at $Z$ and $AX$ is tangent to $\omega$ at $X$).
    \item $B$ is the intersection of the line perpendicular to $OZ$ at $Z$ with the line perpendicular to $OW$ at $W$.
    \item $C$ is the intersection of the line perpendicular to $OY$ at $Y$ with the line perpendicular to $OW$ at $W$.
    \item $D$ is the intersection of the line perpendicular to $OX$ at $X$ and the line perpendicular to $OY$ at $Y$.
    \item $I_1$ is the incenter of the triangle $ABC$.
    \item $I_2$ is the incenter of the triangle $ACD$.
    \item $F_1$ is the orthogonal projection of $I_1$ onto the side $AC$ (so $\omega_1$ has center $I_1$ and goes through $F_1$).
    \item $F_2$ is the orthogonal projection of $I_2$ onto the side $AC$ (so $\omega_2$ has center $I_2$ and goes through $F_2$).
    \item $Q$ is the point of tangency to the circle $\omega_1$ and $T$ is the point of tangent to the circle $\omega_2$ of a common tangent to both cicles.
    \item $P$ is the point of tangency to the circle $\omega_1$ and $S$ is the point of tangency to the circle $\omega_2$ of a common tangent to both circles.
    \item $K$ is the intersection of lines $QT$ and $PS$.
\end{itemize}
Then, prove that $OK=OX$.

\paragraph{Formal translation of the problem:} \

\begin{lstlisting}[language=jgex]
x@4.96_-0.13 y@-1.0068968328888160_-1.2534881080682770 z@-2.8402847238575120_-4.9117762734006830 = triangle x y z; 
o = circle o x y z; 
w@6.9090049230038776_-1.3884003936987552 = on_circle w o x; 
a = on_tline a z o z, on_tline a x o x;
b = on_tline b z o z, on_tline b w o w;
c = on_tline c y o y, on_tline c w o w; 
d = on_tline d x o x, on_tline d y o y; 
i1 = incenter i1 a b c; 
i2 = incenter i2 a c d; 
f1 = foot f1 i1 a c; 
f2 = foot f2 i2 a c; 
q t p s = cc_tangent q t p s i1 f1 i2 f2; 
k = on_line k q t, on_line k p s ? cong o k o x
\end{lstlisting}

\begin{figure}
    \centering
    \includegraphics[height=10cm,width=12cm]{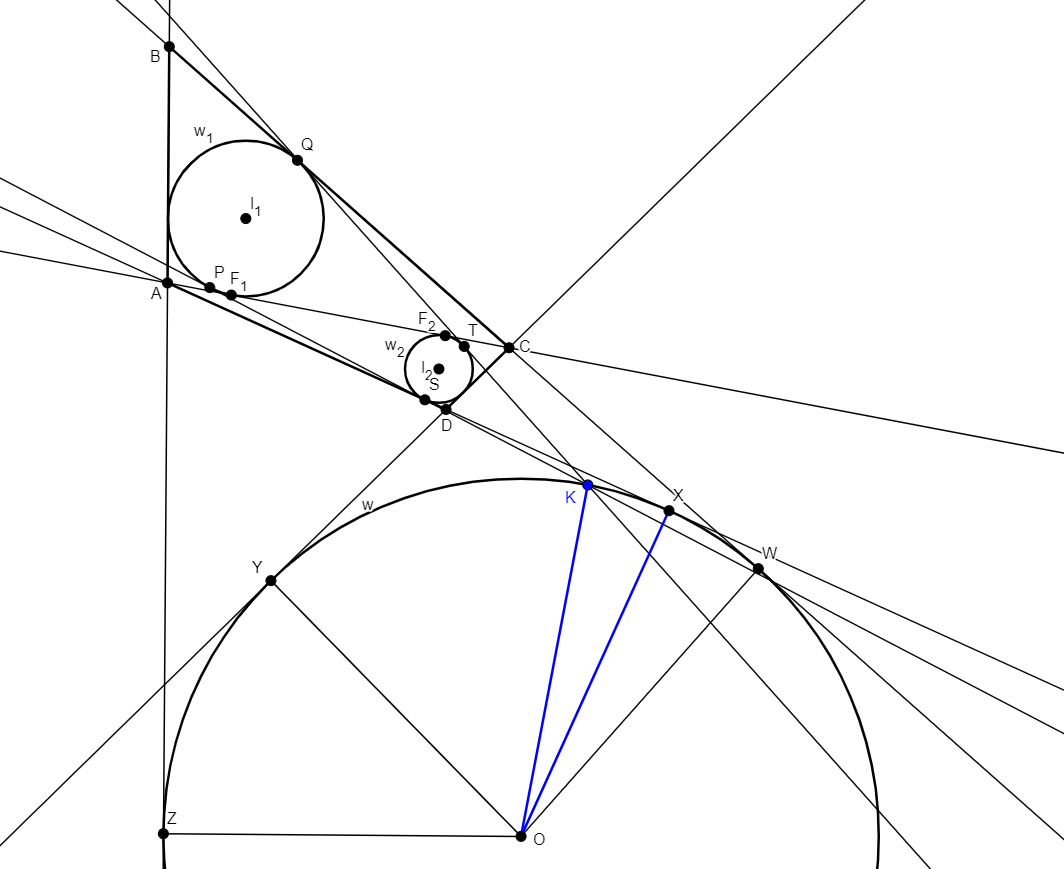}
    \caption{Diagram for the statement of Problem 6 of the IMO 2008 exam.}
\end{figure}

\paragraph{Discussion:} Problem 6 of the 2008 IMO exam is an example of a problem where, for technical reasons, not all the premises given in the problem can be well described by the symbolic system used in DDAR, and that results in an ambiguity that cannot be solved within the design choices of the pseudo-axiomatic system used by AlphaGeometry, and there is no simple way to fix those limitations.

The problem presents two circles, $W1$ and $W2$, and asks something about $K$, the intersection of the common {\textit{external}} tangents to $W1$ and $W2$ specifically, that it lies on a third circle $W$. The issue here is that there are four common tangents to the pair of circles, which results in 6 intersections of pairs of common tangents to the circles. Of those, only one of them lies on $W$.

For a human, it is not hard to make this distinction and find good arguments that fit the correct drawing and that lead towards proof. But this involves a collection of unsaid assumptions about that specific pair of lines that are not available for the computer. Even if the construction created in AlphaGeometry to represent a situation like this (definition \texttt{cc\_tangent}) generates a background coordinate representation of the problem that effectively assumes the choice of the external tangents, there is no symbolical information in terms of predicates that separates the \enquote{good} choice of the pair of tangents from the \enquote{bad} choices. In practical terms, this means that if we could write a proof for this problem, each step of it would apply as well to the other five points that are intersections of common tangents to $W1$ and $W2$. But those do not lie in $W$, and the proof would be false.

Trying to update the foundations to encompass a solution for this problem would involve first creating predicates that would allow the distinction between the different line configurations and then assuring that we have a powerful set of deduction rules that allows for consequences of such predicates to be derived up to the desired results. As the current version of both DDAR and DDARN are not equipped to prescribe and recognize in-betweenness, we envision it will be difficult to devise an axiomatic system that could deal with this problem.

As it remains now, 2008 P6 is a case of undecidability of DDAR as well as DDARN.

\subsection{IMO 2011 P6}

\begin{problem}[Original Formulation of P6 2011 (IMO Shortlist File - G8)]

Let $ABC$ be an acute triangle with circumcircle $\omega$. Let $t$ be a tangent line to $\omega$. Let $t_a$, $t_b$, and $t_c$ be the lines obtained by reflecting $t$ in the lines $BC$, $CA$, and $AB$, respectively. Show that the circumcircle of the triangle determined by the lines $t_a$, $t_b$, and $t_c$ is tangent to the circle $\omega$.
\end{problem}

\paragraph{Explanation of AlphaGeometry's translation of P6 2011:} Consider triangle $ABC$, and also that
\begin{itemize}
    \item $O$ is the circumcenter of $ABC$.
    \item $P$ is a point on the circle centered in $O$ through $A$.
    \item $Q$ is a point on the line perpendicular to $OP$ at $P$ (the tangent line to the circle centered at $O$ at $P$). So the line $PQ$ is $t$.
    \item $P_a$ is the image of $P$ reflected through $BC$.
    \item $P_b$ is the image of $P$ reflected through $CA$.
    \item $P_c$ is the image of $P$ reflected through $AB$.
    \item $Q_a$ is the image of $Q$ reflected through $BC$ (so the line $P_aQ_a$ is $t_a$).
    \item $Q_b$ is the image of $Q$ reflected through $CA$ (so the line $P_bQ_b$ is $t_b$).
    \item $Q_c$ is the image of $Q$ reflected through $AB$ (so the line $P_cQ_c$ is $t_c$).
    \item $A_1$ is the intersection of $t_b$ and $t_c$.
    \item $B_1$ is the intersection of $t_b$ and $t_c$.
    \item $C_1$ is the intersection of $t_a$ and $t_b$.
    \item $O_1$ is the circumcenter of $A_1B_1C_1$.
    \item $X$ is a point in the intersection of the circles of center $O$ through $A$ and of center $O_1$ through $A_1$.
\end{itemize}
Then, prove that $X$, $O$, and $O_1$ are collinear (if the centers of two circles and an intersection point of theirs are collinear, the intersection point is a tangency point of the pair of circles).

\paragraph{Formal translation of our problem:} \

\begin{lstlisting}[language=jgex]
a b c = triangle a b c;
o = circle o a b c;
p = on_circle p o a;
q = on_tline q p o p;
pa = reflect pa p b c;
pb = reflect pb p c a;
pc = reflect pc p a b;
qa = reflect qa q b c;
qb = reflect qb q c a;
qc = reflect qc q a b;
a1 = on_line a1 pb qb, on_line a1 pc qc;
b1 = on_line b1 pa qa, on_line b1 pc qc;
c1 = on_line c1 pa qa, on_line c1 pb qb;
o1 = circle o1 a1 b1 c1;
x = on_circle x o a, on_circle x o1 a1 ? coll x o o1
\end{lstlisting}

\begin{figure}
    \centering
    \includegraphics[height=9cm,width=12cm]{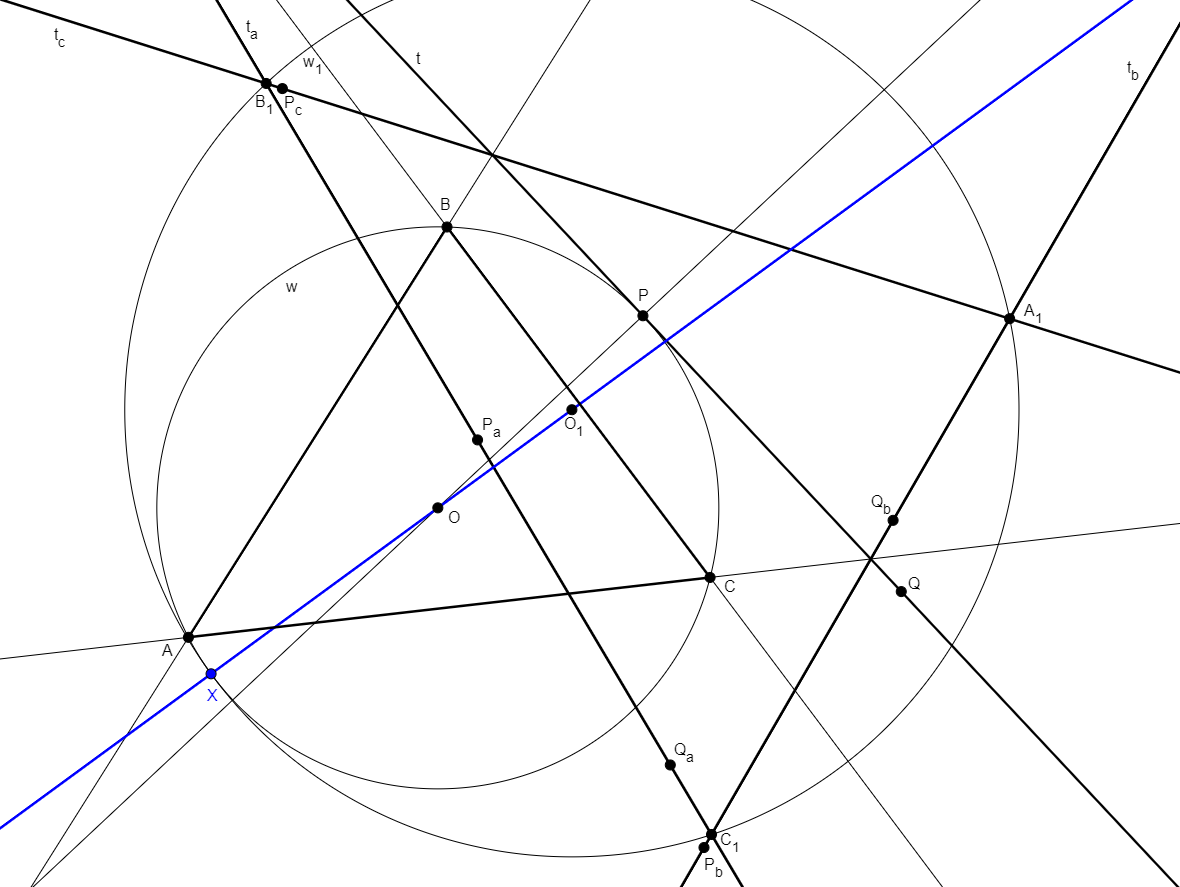}
    \caption{Diagram for the statement of Problem 6 of the IMO 2011 exam.}
\end{figure}

\paragraph{Discussion:} After many attempts to solve this problem through the addition of extra points, no clear candidate for a new minor fix of the engine proved to be a solution. Neither there seems to be a good reason for it to be undecidable. The only commentary that can be made after a long effort is that there seems to be a lack of links between the information associated with point $X$, defined as the intersection of the two circumcircles, and the rest of the problem.

In fact, the proof suggested in the 2011 IMO shortlist file, see~\citep{smit_imo2011s1}, uses the definition of an alternative point, for example, a point $K$, the intersection of another pair of circles, and proceeds to prove $K=X$. This is essentially a proof by contradiction (two distinct points exist then prove that they are the same), a scheme that is not available for AlphaGeometry for two reasons: first, there are checks to prevent the instantiation of two overlapping points, and second, there is no set of axioms that can perform a logical chain ending in \enquote{two points are the same}. It is not impossible to imagine that adaptations could be made to \geosolver{} to allow this sort of argument, but that would take us too far from the scope chosen for this project.

\subsection{IMO 2019 P2}

This problem had a solution presented in the original paper~\citep{trinh_solving_2024}, given by the manual provision of three extra points (which the LLM module could not find), after which the original formulation of DDAR can write a proof.

The human-aided solution to this problem was provided by the authors in their Supplementary Material, and a conjecture on why the transformer could not detect the suggested points in the Extended Data Fig. 4 of the original paper.

\subsection{IMO 2021 P3}

\begin{problem}[Original Formulation of P3 2021 (Evan Chen's Solution Notes)]

Let $D$ be an interior point of the acute triangle $ABC$ with $AB>AC$ so that $\angle DAB=\angle CAD$. The point $E$ on the segment $AC$ satisfies $\angle FDA=\angle DBC$, and the point $X$ on the line $AC$ satisfies $CX=BX$. Let $O_1$ and $O_2$ be the circumcenters of the triangles $ADC$ and $EXD$, respectively. Prove that the lines $BC$, $EF$, and $O_1O_2$ are concurrent.
\end{problem}

\paragraph{Explanation of AlphaGeometry's translation of P3 2021:} Consider triangle $ABC$, and also
\begin{itemize}
    \item $D$ a point in the bisector of angle $\angle BAC$.
    \item $E$ a point on line $AC$ such that $\angle EDA=\angle DCB$.
    \item $F$ a point on line $AB$ such that $\angle FDA=\angle DBC$.
    \item $X$ a point on the intersection of the bisector of segment $BC$ with line $AC$.
    \item $O_1$ the center of the circle through $A$, $D$, and $C$.
    \item $O_2$ the center of the circle through $E$, $X$, and $D$.
    \item $Y$ the intersection of lines $EF$ and $BC$.
\end{itemize}
Then, prove that $O_1$, $O_2$, and $Y$ are collinear.

\paragraph{Formal translation of the problem:} \

\begin{lstlisting}[language=jgex]
a b c = triangle a b c;
d = angle_bisector b a c;
e = on_aline d a d c b, on_line a c;
f = on_aline d a d b c, on_line a b;
x = on_bline b c, on_line a c;
o1 = circle a d c;
o2 = circle e x d;
y = on_line e f, on_line b c ? coll o1 o2 y
\end{lstlisting}

\begin{figure}
    \centering
    \includegraphics[height=8cm,width=12cm]{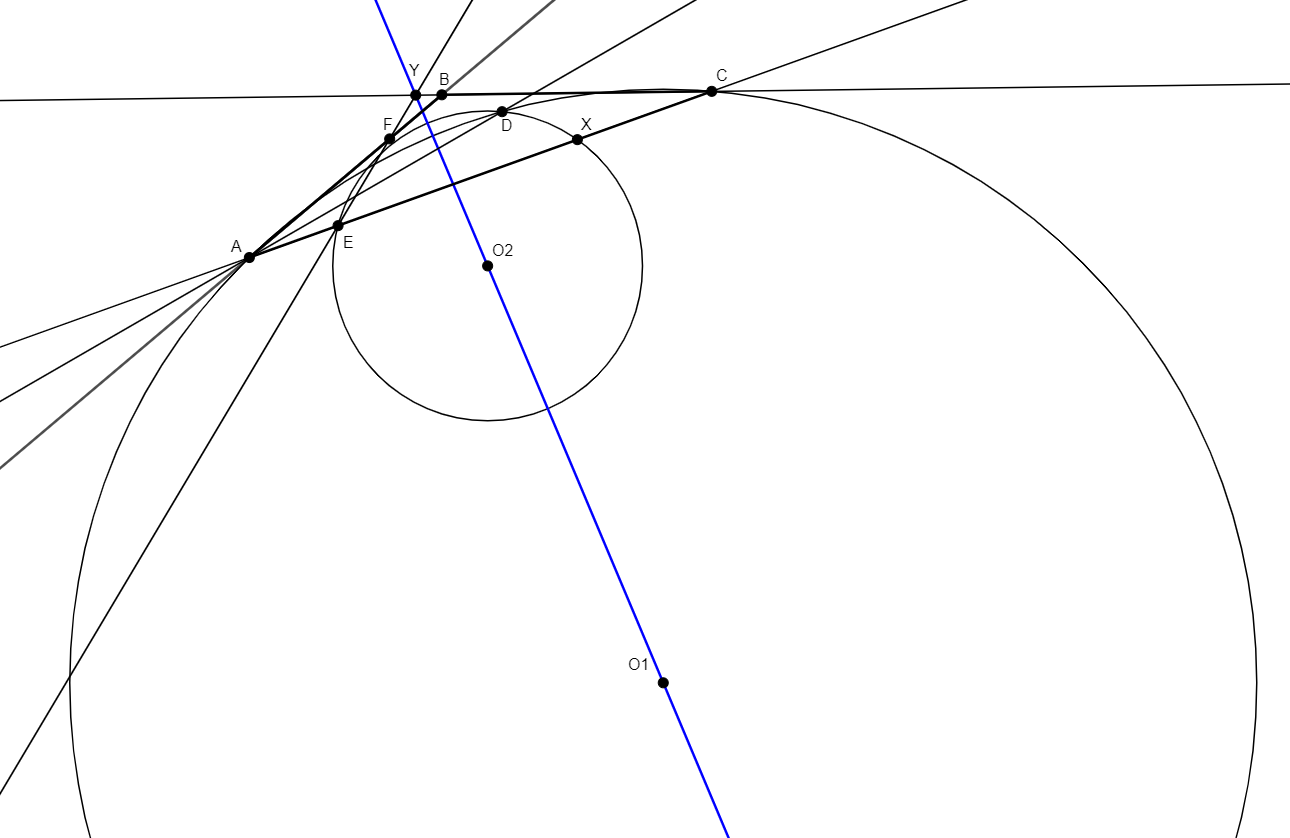}
    \caption{Diagram for the statement of Problem 3 of the IMO 2021 exam.}
\end{figure}

\paragraph{Discussion:} This problem, as 2008 P6, presents a premise that cannot be expressed by AlphaGeometry's engine, namely that point $D$ is interior to the triangle. In general, those kinds of open premises on geometry problems are usually associated with non-degeneracy conditions for the construction of the points of the problem, a situation AlphaGeometry can handle by creating an explicit instance of the problem before starting the solution and checking numerically for the expected goal. If it fails at that point, it will try again with some random variation within its degrees of freedom in the hope that, eventually, it will find a good regime for the construction. Another possible reason for adding open conditions is to avoid a multiplicity of cases that would take too much unnecessary time for the person solving the problem, a consideration AlphaGeometry does not have to deal with.

The case of problem 3 of the 2021 IMO exam is different. Here, if the non-degeneracy condition is violated, we do not have the construction breaking down, but rather, if $D$ is external to $ABC$, the goal {\it{becomes}} false, while all the conditions of the problem remain in place. This means AlphaGeometry will be able to build an instance of the problem and try to proceed to the proof, but as in the case of problem 2008 P6, there is no symbolic difference within AlphaGeometry's symbolic engine between the regime where the goal of the problem is true and where it is false. As a consequence, if a proof were written down for the regime where the problem is true, it would also hold for the regime where the problem is false, hence it would be incorrect. It is also a case of undecidability of the axiomatic system.

Again, to be able to handle such problems is not simply a matter of designing definitions that allow the insertion of information like \enquote{$D$ internal to $ABC$} into the engine. A whole collection of deduction rules must be developed to allow the consequences of that information to be transmitted through the reasoning.

\section{Limitations and Future Work}
\label{sec:user_features}

\paragraph{GeoGebra input} 
There are a couple of drawbacks related to using the GeoGebra constructor. 

First, one has to understand how to perform the geometric construction of the problem with a straightedge and compass sense. Nonetheless, this is also often the case when writing problems in formal language, as the order of the points constructed often has to be coherent with the geometric dependency of some points on others. 

Second, because of the flexibility of calling all numerical tools available to Python when writing a definition, there is a greater flexibility of definitions when proposing a problem with the formal language when compared to the closed set of tools available in GeoGebra. That said, we do believe the ease of use surpasses such drawbacks, especially as the vast majority of problems in plane geometry should admit a GeoGebra construction.

\paragraph{\geosolver's interface.} Currently, \geosolver's user interface is very rudimentary, albeit straightforward. For a normal user, \geosolver{} can import a geometric construction from a GeoGebra file and solve the problem via DDARN, which runs using breadth-first-search or using a Human Agent. 

We plan to improve the interface of \geosolver{} to complete it as a public training gym for auxiliary points finding using AI approaches (including, but not limited to, an LLM-based approach, such as the one from AlphaGeometry), with both text description and geometrical graphics as input and auxiliary points as output. 

\paragraph{\geosolver's LLM module.} Our work to date focused on improving DDAR, which resulted in the new system DDARN. Our motivation for this choice, as argued in the introduction, was that most of the performance benefits of AlphaGeometry come from DDAR.
AlphaGeometry's LLM is written in Meliad. We plan to release a PyTorch version of the LLM, with the same weights as the original model, so that the LLM can be easily integrated and used in the \geosolver{} interface.

\section{Conclusion}
We have presented an improved and expanded version of AlphaGeometry, called \geosolver, that enables user-friendly interaction to solve IMO-level geometry and an agentic interface that can easily be augmented by various agents to guide how DDARN solves problems. Further, we present a much-improved codebase that has better debugging capabilities, and we encourage the community to improve and extend our codebase to enable more user-friendly interactions.

\section*{Acknowledgments}
We thank Ronald Cardenas Acosta for their helpful support during the development of this project.

\appendix

\section{List of rules}
\label{app:rules}

In this section, we have the content of the \texttt{unabridged\_rules.txt} file, containing the largest collection of rules were that used and experimented with in \geosolver. The current implementation of the \texttt{rules.txt} file contains \texttt{r01}, \texttt{r03}-\texttt{r07}, \texttt{r11}-\texttt{r29}, \texttt{r34}, \texttt{r3}, \texttt{r42}, \texttt{r49}, and \texttt{r51}-\texttt{r63} and represents the rules that are used by DDARN. The missing numbers in the \texttt{rules.txt} file correspond to those rules of DDAR, listed in~\citep{trinh_solving_2024} that contain the now unsupported predicates \texttt{simtri*} and \texttt{contri*}. They were replaced by new rules that make the logical steps of the triangle similarity arguments fully explicit. More information on the history of each rule can be found in the relevant documentation section.\footnote{\url{https://lmcrc.github.io/Newclid/manual/default_files/index.html}}

\paragraph{List of rules with names:} \ 

\begin{lstlisting}[language=jgex]
r00 Perpendiculars give parallel
perp A B C D, perp C D E F, ncoll A B E => para A B E F

r01 Definition of circle
cong O A O B, cong O B O C, cong O C O D => cyclic A B C D

r02 Parallel from inclination
eqangle A B P Q C D P Q => para A B C D

r03 Arc determines internal angles
cyclic A B P Q => eqangle P A P B Q A Q B

r04 Congruent angles are in a circle
eqangle P A P B Q A Q B, ncoll P Q A B => cyclic A B P Q

r05 Same arc same chord
cyclic A B C P Q R, eqangle C A C B R P R Q => cong A B P Q

r06 Base of half triangle
midp E A B, midp F A C => para E F B C

r07 Thales Theorem I
para A B C D, coll O A C, coll O B D => eqratio3 A B C D O O

r08 Right triangles common angle I
perp A B C D, perp E F G H, npara A B E F => eqangle A B E F C D G H

r09 Sum of angles of a triangle
eqangle a b c d m n p q, eqangle c d e f p q r u => eqangle a b e f m n r u

r10 Ratio cancellation
eqratio a b c d m n p q, eqratio c d e f p q r u => eqratio a b e f m n r u

r11 Bisector theorem I
eqratio d b d c a b a c, coll d b c, ncoll a b c => eqangle a b a d a d a c

r12 Bisector theorem II
eqangle a b a d a d a c, coll d b c, ncoll a b c => eqratio d b d c a b a c

r13 Isosceles triangle equal angles
cong O A O B, ncoll O A B => eqangle O A A B A B O B

r14 Equal base angles imply isosceles
eqangle A O A B B A B O, ncoll O A B => cong O A O B

r15 Arc determines inscribed angles (tangent)
circle O A B C, perp O A A X => eqangle A X A B C A C B

r16 Same arc giving tangent
circle O A B C, eqangle A X A B C A C B => perp O A A X

r17 Central angle vs inscribed angle I
circle O A B C, midp M B C => eqangle A B A C O B O M

r18 Central angle vs inscribed angle II
circle O A B C, coll M B C, eqangle A B A C O B O M => midp M B C

r19 Hypothenuse is diameter
perp A B B C, midp M A C => cong A M B M

r20 Diameter is hypotenuse
circle O A B C, coll O A C => perp A B B C

r21 Cyclic trapezoid
cyclic A B C D, para A B C D => eqangle A D C D C D C B

r22 Bisector Construction
midp M A B, perp O M A B => cong O A O B

r23 Bisector is perpendicular
cong A P B P, cong A Q B Q => perp A B P Q

r24 Cyclic kite
cong A P B P, cong A Q B Q, cyclic A B P Q => perp P A A Q

r25 Diagonals of parallelogram I
midp M A B, midp M C D => para A C B D

r26 Diagonals of parallelogram II
midp M A B, para A C B D, para A D B C => midp M C D

r27 Thales theorem II
eqratio O A A C O B B D, coll O A C, coll O B D, ncoll A B C, sameside A O C B O D => para A B C D

r28 Overlapping parallels
para A B A C => coll A B C

r29 Midpoint is an eqratio
midp M A B, midp N C D => eqratio M A A B N C C D

r30 Right triangles common angle II
eqangle A B P Q C D U V, perp P Q U V => perp A B C D

r31 Denominator cancelling
eqratio A B P Q C D U V, cong P Q U V => cong A B C D

r34 AA Similarity of triangles (direct)
eqangle B A B C Q P Q R, eqangle C A C B R P R Q, ncoll A B C, sameclock A B C P Q R => simtri A B C P Q R

r35 AA Similarity of triangles (reverse)
eqangle B A B C Q R Q P, eqangle C A C B R Q R P, ncoll A B C, sameclock A B C P R Q => simtrir A B C P Q R

r36 ASA Congruence of triangles (direct)
eqangle B A B C Q P Q R, eqangle C A C B R P R Q, ncoll A B C, cong A B P Q, sameclock A B C P Q R => contri A B C P Q R

r37 ASA Congruence of triangles (reverse)
eqangle B A B C Q R Q P, eqangle C A C B R Q R P, ncoll A B C, cong A B P Q, sameclock A B C P R Q => contrir A B C P Q R

r41 Thales theorem III
para a b c d, coll m a d, coll n b c, eqratio m a m d n b n c, sameside m a d n b c => para m n a b

r42 Thales theorem IV
para a b c d, coll m a d, coll n b c, para m n a b => eqratio m a m d n b n c

r43 Orthocenter theorem
perp a b c d, perp a c b d => perp a d b c

r44 Pappus's theorem
coll a b c, coll p q r, coll x a q, coll x p b, coll y a r, coll y p c, coll z b r, coll z c q => coll x y z

r45 Simson's line theorem
cyclic a b c p, coll a l c, perp p l a c, coll m b c, perp p m b c, coll n a b, perp p n a b => coll l m n

r46 Incenter theorem
eqangle a b a x a x a c, eqangle b a b x b x b c, ncoll a b c => eqangle c b c x c x c a

r47 Circumcenter theorem
midp m a b, perp x m a b, midp n b c, perp x n b c, midp p c a => perp x p c a

r48 Centroid theorem
midp m a b, coll m x c, midp n b c, coll n x a, midp p c a => coll x p b

r49 Recognize center of cyclic (circle)
circle O A B C, cyclic A B C D => cong O A O D

r50 Recognize center of cyclic (cong)
cyclic A B C D, cong O A O B, cong O C O D, npara A B C D => cong O A O C

r51 Midpoint splits in two
midp M A B => rconst M A A B 1/2

r52 Properties of similar triangles (Direct)
simtri A B C P Q R => eqangle B A B C Q P Q R, eqratio B A B C Q P Q R

r53 Properties of similar triangles (Reverse)
simtrir A B C P Q R => eqangle B A B C Q R Q P, eqratio B A B C Q P Q R

r54 Definition of midpoint
cong M A M B, coll M A B => midp M A B

r55 Properties of midpoint (cong)
midp M A B => cong M A M B

r56 Properties of midpoint (coll)
midp M A B => coll M A B

r57 Pythagoras theorem
PythagoreanPremises a b c => PythagoreanConclusions a b c

r58 Same chord same arc I
cyclic a b c p q r, cong a b p q, sameclock c a b r p q, sameside c a b r p q => eqangle c a c b r p r q

r59 Same chord same arc II
cyclic a b c p q r, cong a b p q, sameclock c b a r p q, nsameside c b a r p q => eqangle c a c b r q r p

r60 SSS Similarity of triangles (Direct)
eqratio B A B C Q P Q R, eqratio C A C B R P R Q, ncoll A B C, sameclock A B C P Q R => simtri A B C P Q R

r61 SSS Similarity of triangles (Reverse)
eqratio B A B C Q P Q R, eqratio C A C B R P R Q, ncoll A B C, sameclock A B C P R Q => simtrir A B C P Q R

r62 SAS Similarity of triangles (Direct)
eqratio B A B C Q P Q R, eqangle B A B C Q P Q R, ncoll A B C, sameclock A B C P Q R => simtri A B C P Q R

r63 SAS Similarity of triangles (Reverse)
eqratio B A B C Q P Q R, eqangle B A B C Q P Q R, ncoll A B C, sameclock A B C P R Q => simtrir A B C P Q R

r64 SSS Congruence of triangles (Direct)
cong A B P Q, cong B C Q R, cong C A R P, ncoll A B C, sameclock A B C P Q R => contri A B C P Q R

r65 SSS Congruence of triangles (Reverse)
cong A B P Q, cong B C Q R, cong C A R P, ncoll A B C, sameclock A B C P R Q => contrir A B C P Q R

r66 SAS Congruence of triangles (Direct)
cong A B P Q, cong B C Q R, eqangle B A B C Q P Q R, ncoll A B C, sameclock A B C P Q R => contri A B C P Q R

r67 SAS Congruence of triangles (Reverse)
cong A B P Q, cong B C Q R, eqangle B A B C Q P Q R, ncoll A B C, sameclock A B C P R Q => contrir A B C P Q R

r68 Similarity without scaling (Direct)
eqratio B A B C Q P Q R, eqratio C A C B R P R Q, ncoll A B C, cong A B P Q, sameclock A B C P Q R => contri A B C P Q R

r69 Similarity without scaling (Reverse)
eqratio B A B C Q P Q R, eqratio C A C B R P R Q, ncoll A B C, cong A B P Q, sameclock A B C P R Q => contrir A B C P Q R
\end{lstlisting}

\pagebreak
\section{Full Solution IMO 2008 P1}
\label{app:full_solution}

Here we present the full proof of problem 1 from the IMO 2008 exam discussed in Subsection~\ref{subsec:imo_2008_p1b}, as generated by \geosolver. The auxiliary constructions were added by a human. DDAR was not able to solve this problem even if provided the extra points.

\begin{lstlisting}[language=jgex]
==========================
 * From theorem premises:
A B C H D E F A1 A2 B1 B2 C1 C2 : Points
AH ⟂ BC [00]
BH ⟂ AC [01]
B,C,D are collinear [02]
DB = DC [03]
A,E,C are collinear [04]
EA = EC [05]
B,A,F are collinear [06]
FA = FB [07]
B,A_1,C are collinear [08]
DA_1 = DH [09]
B,A_2,C are collinear [10]
DA_2 = DH [11]
A,B_1,C are collinear [12]
EB_1 = EH [13]
A,C,B_2 are collinear [14]
EB_2 = EH [15]
B,A,C_1 are collinear [16]
FC_1 = FH [17]
B,A,C_2 are collinear [18]
FC_2 = FH [19]

 * Auxiliary Constructions:
O1 O2 O3 O : Points
O_1E = EH [20]
O_1H ⟂ EF [21]
O_2E = EH [22]
DE ⟂ HO_2 [23]
O_3F = FH [24]
HO_3 ⟂ DF [25]
OC_1 = OC_2 [26]
OC_2 = OB_1 [27]

 * Proof steps:
001. OC_1 = OC_2 [26] & OC_2 = OB_1 [27] (why_cong_resolution)⇒  OB_1 = OC_1 [28]
002. C_1O = B_1O [28] & OC_1 = OC_2 [26] (why_circle_resolution)⇒  O is the circumcenter of \Delta C_1B_1C_2 [29]
003. A,C,B_2 are collinear [14] & A,C,B_1 are collinear [12] (i04)⇒  A,B_1,B_2 are collinear [30]
004. B,A,C_2 are collinear [18] & B,A,C_1 are collinear [16] (i04)⇒  B,A,C_2,C_1 are collinear [31]
005. EB_1 = EH [13] & O_1E = EH [20] & EB_2 = EH [15] (i01)⇒  O_1,B_1,B_2,H are concyclic [32]
006. O_1,B_1,B_2,H are concyclic [32] (r03)⇒  ∠O_1B_2B_1 = ∠O_1HB_1 [33]
007. ∠O_1B_2B_1 = ∠O_1HB_1 [33] & A,B_1,B_2 are collinear [30] (i08)⇒  ∠(AB_1-B_2O_1) = ∠B_1HO_1 [34]
008. ∠O_1B_2B_1 = ∠O_1HB_1 [33] & A,B_1,B_2 are collinear [30] (i08)⇒  ∠HO_1B_2 = ∠HB_1A [35]
009. B,A_2,C are collinear [10] & B,C,A_1 are collinear [08] (i04)⇒  B,A_1,A_2 are collinear [36]
010. B,A_2,C are collinear [10] & B,A_2,A_1 are collinear [36] (i04)⇒  B,A_1,C,A_2 are collinear [37]
011. AH ⟂ BC [00] ()⇒  ∠(AH-BC) = ∠(BC-AH) [38]
012. B,A_1,C are collinear [08] ()⇒  BC ∥ A_1B [39]
013. ∠(AH-BC) = ∠(BC-AH) [38] & BC ∥ A_1B [39] (_why_perp_repr)⇒  AH ⟂ A_1B [40]
014. B,A_1,A_2,C are collinear [37] & AH ⟂ A_1B [40] (why_perp_resolution)⇒  AH ⟂ A_2C [41]
015. O_1H ⟂ EF [21] & AH ⟂ A_2C [41] (r08)⇒  ∠O_1HA = ∠(EF-A_2C) [42]
016. ∠O_1HA = ∠(EF-A_2C) [42] & B,A_1,C,A_2 are collinear [37] (i08)⇒  ∠(EF-A_1B) = ∠O_1HA [43]
017. F,A,B are collinear [06] & FA = FB [07] (why_midp_resolution)⇒  F is midpoint of AB [44]
018. E,A,C are collinear [04] & EA = EC [05] (why_midp_resolution)⇒  E is midpoint of AC [45]
019. F is midpoint of AB [44] & E is midpoint of AC [45] (r06)⇒  FE ∥ BC [46]
020. FE ∥ BC [46] & B,A_1,C are collinear [08] (i05)⇒  EF ∥ A_1B [47]
021. ∠(EF-A_1B) = ∠O_1HA [43] & EF ∥ A_1B [47] (i03)⇒  HO_1 ∥ AH [48]
022. ∠(AB_1-B_2O_1) = ∠B_1HO_1 [34] & HO_1 ∥ AH [48] (_why_eqangle_eqangle)⇒  ∠(AB_1-B_2O_1) = ∠B_1HA [49]
023. A,B_2,B_1 are collinear [30] & ∠(AB_1-B_2O_1) = ∠B_1HA [49] (why_eqangle6_resolution)⇒  ∠AB_2O_1 = ∠B_1HA [50]
024. HA ∥ HO_1 [48] (r28)⇒  A,H,O_1 are collinear [51]
025. ∠HO_1B_2 = ∠HB_1A [35] & HO_1 ∥ AH [48] (_why_eqangle_eqangle)⇒  ∠(AH-B_2O_1) = ∠HB_1A [52]
026. O_1,H,A are collinear [51] & ∠(AH-B_2O_1) = ∠HB_1A [52] (why_eqangle6_resolution)⇒  ∠AO_1B_2 = ∠HB_1A [53]
027. ∠AB_2O_1 = ∠B_1HA [50] & ∠AO_1B_2 = ∠HB_1A [53] (Similar Triangles 35)⇒  AB_2:AO_1 = AH:AB_1 [54]
028. O_2E = EH [22] & EB_1 = EH [13] & EB_2 = EH [15] (why_circle_resolution)⇒  E is the circumcenter of \Delta B_1O_2B_2 [55]
029. O_2E = EH [22] & EB_1 = EH [13] & EB_2 = EH [15] (i01)⇒  B_2,B_1,H,O_2 are concyclic [56]
030. A,C,B_2 are collinear [14] & A,C,B_1 are collinear [12] & A,C,E are collinear [04] (i04)⇒  A,B_1,B_2,E are collinear [57]
031. A,B_1,B_2,E are collinear [57] (why_coll_resolution)⇒  E,B_1,B_2 are collinear [58]
032. E is the circumcenter of \Delta B_1O_2B_2 [55] & E,B_1,B_2 are collinear [58] (r20)⇒  B_1O_2 ⟂ B_2O_2 [59]
033. B_2,B_1,H,O_2 are concyclic [56] (r03)⇒  ∠B_1HB_2 = ∠B_1O_2B_2 [60]
034. B_2,B_1,H,O_2 are concyclic [56] (r03)⇒  ∠B_2HO_2 = ∠B_2B_1O_2 [61]
035. AH ⟂ BC [00] & B_1O_2 ⟂ B_2O_2 [59] & ∠B_1HB_2 = ∠B_1O_2B_2 [60] ()⇒  ∠B_1HB_2 = ∠(BC-AH) [62]
036. ∠B_1HB_2 = ∠(BC-AH) [62] & BC ∥ A_1B [39] (_why_eqangle_eqangle)⇒  ∠B_1HB_2 = ∠(A_1B-AH) [63]
037. B,A_1,A_2,C are collinear [37] & ∠B_1HB_2 = ∠(A_1B-AH) [63] (why_eqangle_resolution)⇒  ∠B_1HB_2 = ∠(A_2C-AH) [64]
038. B_1,B_2,H,O_1 are concyclic [32] & O_2,B_1,B_2,H are concyclic [56] (why_cyclic_resolution)⇒  B_2,B_1,O_2,O_1 are concyclic [65]
039. B_2,B_1,O_2,H are concyclic [56] & B_2,B_1,O_2,O_1 are concyclic [65] (i11)⇒  B_1,B_2,O_2,O_1,H are concyclic [66]
040. B_1,B_2,O_2,O_1,H are concyclic [66] (why_cyclic_resolution)⇒  O_1,B_2,H,O_2 are concyclic [67]
041. B_1,B_2,O_2,O_1,H are concyclic [66] (why_cyclic_resolution)⇒  O_1,O_2,H,B_1 are concyclic [68]
042. O_1,B_2,H,O_2 are concyclic [67] (r03)⇒  ∠HO_1O_2 = ∠HB_2O_2 [69]
043. ∠HB_2O_2 = ∠HO_1O_2 [69] & HO_1 ∥ AH [48] (why_eqangle_resolution)⇒  ∠HB_2O_2 = ∠(AH-O_1O_2) [70]
044. ∠B_1HB_2 = ∠(A_2C-AH) [64] & ∠HB_2O_2 = ∠(AH-O_1O_2) [70] (r09)⇒  ∠(HB_1-B_2O_2) = ∠(A_2C-O_1O_2) [71]
045. ∠(HB_1-B_2O_2) = ∠(A_2C-O_1O_2) [71] & B,A_1,C,A_2 are collinear [37] (i08)⇒  ∠(B_1H-B_2O_2) = ∠(A_1B-O_1O_2) [72]
046. ∠(B_1H-B_2O_2) = ∠(A_1B-O_1O_2) [72] & EF ∥ A_1B [47] (i09)⇒  ∠(B_1H-B_2O_2) = ∠(EF-O_1O_2) [73]
047. ∠B_1HB_2 = ∠B_1O_2B_2 [60] & B_1O_2 ⟂ B_2O_2 [59] (why_eqangle_resolution)⇒  ∠B_2O_2B_1 = ∠B_1HB_2 [74]
048. EH = EB_2 [15] (r13)⇒  ∠EHB_2 = ∠HB_2E [75]
049. ∠EHB_2 = ∠HB_2E [75] & A,B_1,B_2,E are collinear [57] (i08)⇒  ∠EHB_2 = ∠(B_2H-AB_1) [76]
050. ∠B_2HO_2 = ∠B_2B_1O_2 [61] & A,B_1,B_2 are collinear [30] (i08)⇒  ∠(AB_1-B_2H) = ∠B_1O_2H [77]
051. ∠B_2HE = ∠(AB_1-B_2H) [76] & ∠(AB_1-B_2H) = ∠B_1O_2H [77] (why_eqangle_resolution)⇒  ∠B_1O_2H = ∠B_2HE [78]
052. ∠B_2O_2B_1 = ∠B_1HB_2 [74] & ∠B_1O_2H = ∠B_2HE [78] (r09)⇒  ∠(B_2O_2-B_1H) = ∠O_2HE [79]
053. ∠(B_1H-B_2O_2) = ∠(EF-O_1O_2) [73] & ∠(B_1H-B_2O_2) = ∠EHO_2 [79] ()⇒  ∠EHO_2 = ∠(EF-O_1O_2) [80]
054. FC_1 = FH [17] & FC_2 = FH [19] (why_cong_resolution)⇒  C_2F = C_1F [81]
055. C_2F = C_1F [81] & C_2O = C_1O [26] (r23)⇒  C_2C_1 ⟂ FO [82]
056. D,C,B are collinear [02] & DB = DC [03] (why_midp_resolution)⇒  D is midpoint of CB [83]
057. D is midpoint of CB [83] & E is midpoint of CA [45] (r06)⇒  DE ∥ AB [84]
058. C_1,B,A,C_2 are collinear [31] & DE ∥ AB [84] (why_para_resolution)⇒  C_2C_1 ∥ DE [85]
059. C_2C_1 ⟂ FO [82] & DE ⟂ HO_2 [23] & C_2C_1 ∥ DE [85] (i00)⇒  FO ∥ HO_2 [86]
060. ∠EHO_2 = ∠(EF-O_1O_2) [80] & HO_2 ∥ FO [86] & EF ∥ A_1B [47] (_why_eqangle_eqangle)⇒  ∠(EH-FO) = ∠(A_1B-O_1O_2) [87]
061. B,A_1,A_2,C are collinear [37] & ∠(EH-FO) = ∠(A_1B-O_1O_2) [87] (why_eqangle_resolution)⇒  ∠(EH-FO) = ∠(A_2C-O_1O_2) [88]
062. O_1,O_2,B_2,B_1 are concyclic [65] (r03)⇒  ∠B_2O_1B_1 = ∠B_2O_2B_1 [89]
063. O_1,O_2,B_2,B_1 are concyclic [65] (r03)⇒  ∠O_1B_1B_2 = ∠O_1O_2B_2 [90]
064. ∠B_2O_1B_1 = ∠B_2O_2B_1 [89] & B_1O_2 ⟂ B_2O_2 [59] (why_eqangle_resolution)⇒  ∠B_1O_2B_2 = ∠B_2O_1B_1 [91]
065. EB_1 = EH [13] & O_1E = EH [20] (why_cong_resolution)⇒  EO_1 = EB_1 [92]
066. EO_1 = EB_1 [92] (r13)⇒  ∠EO_1B_1 = ∠O_1B_1E [93]
067. A,C,B_1 are collinear [12] & A,C,E are collinear [04] (i04)⇒  A,B_1,E are collinear [94]
068. ∠EO_1B_1 = ∠O_1B_1E [93] & A,B_1,E are collinear [94] (i08)⇒  ∠EO_1B_1 = ∠O_1B_1A [95]
069. ∠O_1B_1B_2 = ∠O_1O_2B_2 [90] & A,B_1,B_2 are collinear [30] (i08)⇒  ∠AB_1O_1 = ∠B_2O_2O_1 [96]
070. ∠B_1O_1E = ∠AB_1O_1 [95] & ∠AB_1O_1 = ∠B_2O_2O_1 [96] (why_eqangle_resolution)⇒  ∠B_2O_2O_1 = ∠B_1O_1E [97]
071. ∠B_1O_2B_2 = ∠B_2O_1B_1 [91] & ∠B_2O_2O_1 = ∠B_1O_1E [97] (r09)⇒  ∠(B_1O_2-B_2O_1) = ∠O_2O_1E [98]
072. AH ⟂ BC [00] & B_1O_2 ⟂ B_2O_2 [59] & ∠B_2O_1B_1 = ∠B_2O_2B_1 [89] ()⇒  ∠B_2O_1B_1 = ∠(BC-AH) [99]
073. ∠B_2O_1B_1 = ∠(BC-AH) [99] & BC ∥ A_1B [39] (_why_eqangle_eqangle)⇒  ∠B_2O_1B_1 = ∠(A_1B-AH) [100]
074. B,A_1,A_2,C are collinear [37] & ∠B_2O_1B_1 = ∠(A_1B-AH) [100] (why_eqangle_resolution)⇒  ∠B_2O_1B_1 = ∠(A_2C-AH) [101]
075. O_1,O_2,H,B_1 are concyclic [68] (r03)⇒  ∠O_1HO_2 = ∠O_1B_1O_2 [102]
076. ∠O_1B_1O_2 = ∠O_1HO_2 [102] & HO_1 ∥ AH [48] (why_eqangle_resolution)⇒  ∠O_1B_1O_2 = ∠AHO_2 [103]
077. ∠B_2O_1B_1 = ∠(A_2C-AH) [101] & ∠O_1B_1O_2 = ∠AHO_2 [103] (r09)⇒  ∠(O_1B_2-B_1O_2) = ∠(A_2C-HO_2) [104]
078. ∠(O_1B_2-B_1O_2) = ∠(A_2C-HO_2) [104] & B,A_1,C,A_2 are collinear [37] (i08)⇒  ∠(B_2O_1-B_1O_2) = ∠(A_1B-HO_2) [105]
079. ∠(B_2O_1-B_1O_2) = ∠(A_1B-HO_2) [105] & EF ∥ A_1B [47] & FO ∥ HO_2 [86] (i09)⇒  ∠(B_1O_2-B_2O_1) = ∠(HO_2-EF) [106]
080. ∠(B_1O_2-B_2O_1) = ∠O_2O_1E [98] & ∠(B_1O_2-B_2O_1) = ∠(HO_2-EF) [106] ()⇒  ∠(HO_2-EF) = ∠O_2O_1E [107]
081. ∠(HO_2-EF) = ∠O_2O_1E [107] & HO_2 ∥ FO [86] & EF ∥ A_1B [47] (_why_eqangle_eqangle)⇒  ∠(FO-A_1B) = ∠O_2O_1E [108]
082. B,A_1,A_2,C are collinear [37] & ∠(FO-A_1B) = ∠O_2O_1E [108] (why_eqangle_resolution)⇒  ∠(FO-A_2C) = ∠O_2O_1E [109]
083. ∠(EH-FO) = ∠(A_2C-O_1O_2) [88] & ∠(FO-A_2C) = ∠O_2O_1E [109] (r09)⇒  ∠(EH-A_2C) = ∠(A_2C-EO_1) [110]
084. ∠(EH-A_2C) = ∠(A_2C-EO_1) [110] & B,A_1,C,A_2 are collinear [37] (i08)⇒  ∠(EH-A_1B) = ∠(A_1B-EO_1) [111]
085. ∠(EH-A_1B) = ∠(A_1B-EO_1) [111] & EF ∥ A_1B [47] (i09)⇒  ∠HEF = ∠FEO_1 [112]
086. HE = O_1E [20] & ∠HEF = ∠FEO_1 [112] (SAS 33)⇒  FH = FO_1 [113]
087. FC_2 = FH [19] & O_3F = FH [24] & FH = FO_1 [113] (i01)⇒  C_2,O_1,H,O_3 are concyclic [114]
088. FC_1 = FH [17] & O_3F = FH [24] & FC_2 = FH [19] (i01)⇒  C_2,H,O_3,C_1 are concyclic [115]
089. FC_1 = FH [17] & O_3F = FH [24] & FC_2 = FH [19] (why_circle_resolution)⇒  F is the circumcenter of \Delta C_2O_3C_1 [116]
090. C_2,O_1,H,O_3 are concyclic [114] & C_2,H,O_3,C_1 are concyclic [115] (i11)⇒  O_1,C_1,C_2,H are concyclic [117]
091. O_1,C_1,C_2,H are concyclic [117] (r03)⇒  ∠O_1C_2C_1 = ∠O_1HC_1 [118]
092. ∠O_1C_2C_1 = ∠O_1HC_1 [118] & B,A,C_2,C_1 are collinear [31] & A,H,O_1 are collinear [51] (i08)⇒  ∠(C_2O_1-AB) = ∠AHC_1 [119]
093. ∠O_1C_2C_1 = ∠O_1HC_1 [118] & B,A,C_2,C_1 are collinear [31] & A,H,O_1 are collinear [51] (i08)⇒  ∠(C_2O_1-AH) = ∠(AB-C_1H) [120]
094. B,A,C_2 are collinear [18] & ∠AHC_1 = ∠(C_2O_1-AB) [119] (why_eqangle6_resolution)⇒  ∠AHC_1 = ∠O_1C_2A [121]
095. C_1,A,B are collinear [16] & O_1,H,A are collinear [51] & ∠(AB-C_1H) = ∠(C_2O_1-AH) [120] (why_eqangle6_resolution)⇒  ∠AC_1H = ∠C_2O_1A [122]
096. ∠AHC_1 = ∠O_1C_2A [121] & ∠AC_1H = ∠C_2O_1A [122] (Similar Triangles 35)⇒  AH:AC_1 = AC_2:AO_1 [123]
097. OC_1 = OC_2 [26] (r13)⇒  ∠OC_1C_2 = ∠C_1C_2O [124]
098. ∠OC_1C_2 = ∠C_1C_2O [124] & B,A,C_2,C_1 are collinear [31] (i08)⇒  ∠(C_1O-AB) = ∠(AB-C_2O) [125]
099. C_1,A,B are collinear [16] & B,A,C_2 are collinear [18] & ∠(C_1O-AB) = ∠(AB-C_2O) [125] (why_eqangle6_resolution)⇒  ∠OC_1A = ∠BC_2O [126]
100. HO_2 ⟂ DE [23] ()⇒  ∠(DE-HO_2) = ∠(HO_2-DE) [127]
101. ∠(DE-HO_2) = ∠(HO_2-DE) [127] & DE ∥ AB [84] & HO_2 ∥ FO [86] (_why_eqangle_eqangle)⇒  ∠(AB-FO) = ∠(FO-AB) [128]
102. A,F,B are collinear [06] & ∠(AB-FO) = ∠(FO-AB) [128] (why_eqangle6_resolution)⇒  ∠BFO = ∠OFA [129]
103. BF = AF [07] & ∠BFO = ∠OFA [129] (SAS 33)⇒  ∠FBO = ∠OAF [130]
104. ∠FBO = ∠OAF [130] & B,A,F are collinear [06] (i08)⇒  ∠ABO = ∠OAB [131]
105. C_1,A,B are collinear [16] & C_2,A,B are collinear [18] & ∠OAB = ∠ABO [131] (why_eqangle6_resolution)⇒  ∠OAC_1 = ∠C_2BO [132]
106. ∠OC_1A = ∠BC_2O [126] & ∠OAC_1 = ∠C_2BO [132] (Similar Triangles 35)⇒  C_1O:C_2O = C_1A:C_2B [133]
107. C_1O:C_2O = C_1A:C_2B [133] & C_1O = C_2O [26] (i02)⇒  C_1A = C_2B [134]
108. AH:AC_1 = AC_2:AO_1 [123] & C_1A = C_2B [134] (i10)⇒  AH:C_2B = AC_2:AO_1 [135]
109. AB_2:AO_1 = AH:AB_1 [54] & AH:C_2B = AC_2:AO_1 [135] (Ratio chase)⇒  AB_1:C_2B = AC_2:AB_2 [136]
110. AC_2:AB_2 = AB_1:C_2B [136] & C_1A = C_2B [134] (why_eqratio6_resolution)⇒  AC_2:AB_2 = AB_1:AC_1 [137]
111. F is midpoint of BA [44] & D is midpoint of BC [83] (r06)⇒  FD ∥ AC [138]
112. FD ∥ AC [138] & A,B_1,C are collinear [12] (i05)⇒  DF ∥ AB_1 [139]
113. DE ∥ AB [84] & DF ∥ AB_1 [139] (_why_eqangle_eqangle)⇒  ∠BAB_1 = ∠BAB_1 [140]
114. B,A,C_2 are collinear [18] & A,B_2,B_1 are collinear [30] & C_1,A,B are collinear [16] & ∠BAB_1 = ∠BAB_1 [140] (why_eqangle6_resolution)⇒  ∠C_2AB_2 = ∠C_1AB_1 [141]
115. AC_2:AB_2 = AB_1:AC_1 [137] & ∠C_2AB_2 = ∠C_1AB_1 [141] (Similar Triangles 39)⇒  ∠AC_2B_2 = ∠C_1B_1A [142]
116. ∠AC_2B_2 = ∠C_1B_1A [142] & B,A,C_2 are collinear [18] (i08)⇒  ∠(AB-B_2C_2) = ∠C_1B_1A [143]
117. A,B_2,B_1 are collinear [30] & C_1,B,A,C_2 are collinear [31] & ∠C_1B_1A = ∠(AB-B_2C_2) [143] (why_eqangle6_resolution)⇒  ∠C_1B_1B_2 = ∠C_1C_2B_2 [144]
118. ∠C_1B_1B_2 = ∠C_1C_2B_2 [144] (r04)⇒  C_1,B_1,C_2,B_2 are concyclic [145]
119. O is the circumcenter of \Delta C_1B_1C_2 [29] & C_1,B_1,C_2,B_2 are concyclic [145] (r49)⇒  OC_1 = OB_2 [146]
120. O is the circumcenter of \Delta C_1B_1C_2 [29] & C_1,B_1,C_2,B_2 are concyclic [145] (r49)⇒  B_1O = B_2O [147]
121. C_1,O_3,H,C_2 are concyclic [115] (r03)⇒  ∠C_1HO_3 = ∠C_1C_2O_3 [148]
122. C_1,O_3,H,C_2 are concyclic [115] (r03)⇒  ∠C_1C_2H = ∠C_1O_3H [149]
123. C_1,O_3,H,C_2 are concyclic [115] (r03)⇒  ∠C_2HC_1 = ∠C_2O_3C_1 [150]
124. ∠C_1HO_3 = ∠C_1C_2O_3 [148] & B,A,C_2,C_1 are collinear [31] (i08)⇒  ∠O_3HC_1 = ∠(C_2O_3-AB) [151]
125. ∠C_1HO_3 = ∠C_1C_2O_3 [148] & B,A,C_2,C_1 are collinear [31] (i08)⇒  ∠(AB-C_1H) = ∠C_2O_3H [152]
126. BH ⟂ AC [01] ()⇒  ∠(BH-AC) = ∠(AC-BH) [153]
127. A,B_1,C are collinear [12] ()⇒  AC ∥ AB_1 [154]
128. ∠(BH-AC) = ∠(AC-BH) [153] & AC ∥ AB_1 [154] (_why_perp_repr)⇒  BH ⟂ AB_1 [155]
129. E,A,B_1 are collinear [94] & BH ⟂ AB_1 [155] (why_perp_resolution)⇒  HB ⟂ EA [156]
130. HB ⟂ EA [156] & HO_3 ⟂ DF [25] (r08)⇒  ∠BHO_3 = ∠(EA-DF) [157]
131. HB ⟂ EA [156] & HO_3 ⟂ DF [25] (r08)⇒  ∠(DF-HB) = ∠(HO_3-EA) [158]
132. ∠BHO_3 = ∠(EA-DF) [157] & A,B_1,E are collinear [94] (i08)⇒  ∠(AB_1-DF) = ∠BHO_3 [159]
133. ∠(AB_1-DF) = ∠BHO_3 [159] & AB_1 ∥ DF [139] (i03)⇒  BH ∥ HO_3 [160]
134. ∠(C_2O_3-AB) = ∠O_3HC_1 [151] & HO_3 ∥ BH [160] (_why_eqangle_eqangle)⇒  ∠(C_2O_3-AB) = ∠BHC_1 [161]
135. B,A,C_2 are collinear [18] & ∠(C_2O_3-AB) = ∠BHC_1 [161] (why_eqangle6_resolution)⇒  ∠O_3C_2B = ∠BHC_1 [162]
136. HO_3 ∥ HB [160] (r28)⇒  O_3,H,B are collinear [163]
137. DE ∥ AB [84] (_why_eqangle_eqangle)⇒  ∠HBA = ∠HBA [164]
138. O_3,H,B are collinear [163] & C_2,A,B are collinear [18] & C_1,A,B are collinear [16] & ∠HBA = ∠HBA [164] (why_eqangle6_resolution)⇒  ∠O_3BC_2 = ∠HBC_1 [165]
139. ∠O_3C_2B = ∠BHC_1 [162] & ∠O_3BC_2 = ∠HBC_1 [165] (Similar Triangles 35)⇒  O_3C_2:O_3B = C_1H:C_1B [166]
140. ∠O_3C_2B = ∠BHC_1 [162] & ∠O_3BC_2 = ∠HBC_1 [165] (Similar Triangles 35)⇒  O_3C_2:C_2B = C_1H:HB [167]
141. C_1,A,B are collinear [16] & B,A,C_2 are collinear [18] & ∠(C_1O-AB) = ∠(AB-C_2O) [125] (why_eqangle6_resolution)⇒  ∠OC_1B = ∠AC_2O [168]
142. C_1,A,B are collinear [16] & B,A,C_2 are collinear [18] & ∠OBA = ∠BAO [131] (why_eqangle6_resolution)⇒  ∠OBC_1 = ∠C_2AO [169]
143. ∠OC_1B = ∠AC_2O [168] & ∠OBC_1 = ∠C_2AO [169] (Similar Triangles 35)⇒  C_1O:C_2O = C_1B:C_2A [170]
144. C_1O:C_2O = C_1B:C_2A [170] & C_1O = C_2O [26] (i02)⇒  C_1B = C_2A [171]
145. O_3C_2:O_3B = C_1H:C_1B [166] & C_1B = C_2A [171] (i10)⇒  O_3C_2:O_3B = C_1H:AC_2 [172]
146. B,A,C_1 are collinear [16] & B,A,F are collinear [06] (i04)⇒  B,A,F,C_1 are collinear [173]
147. B,A,F,C_1 are collinear [173] & B,A,C_2 are collinear [18] (why_coll_resolution)⇒  F,C_2,C_1 are collinear [174]
148. F is the circumcenter of \Delta C_2O_3C_1 [116] & F,C_2,C_1 are collinear [174] (r20)⇒  C_2O_3 ⟂ C_1O_3 [175]
149. C_2O_3 ⟂ C_1O_3 [175] & HO_3 ⟂ DF [25] (why_eqangle_resolution)⇒  ∠(DF-HO_3) = ∠C_2O_3C_1 [176]
150. C_2O_3 ⟂ C_1O_3 [175] & HO_3 ⟂ DF [25] (r08)⇒  ∠(DF-C_2O_3) = ∠HO_3C_1 [177]
151. ∠C_1C_2H = ∠C_1O_3H [149] & B,A,C_2,C_1 are collinear [31] (i08)⇒  ∠(C_1O_3-AB) = ∠O_3HC_2 [178]
152. A,F,B are collinear [06] & ∠O_3HC_2 = ∠(C_1O_3-AB) [178] (why_eqangle_resolution)⇒  ∠O_3HC_2 = ∠(C_1O_3-AF) [179]
153. ∠(DF-HO_3) = ∠C_2O_3C_1 [176] & ∠O_3HC_2 = ∠(C_1O_3-AF) [179] (r09)⇒  ∠(DF-HC_2) = ∠(O_3C_2-AF) [180]
154. ∠(DF-HC_2) = ∠(O_3C_2-AF) [180] & B,A,F are collinear [06] (i08)⇒  ∠(DF-C_2H) = ∠(C_2O_3-AB) [181]
155. O_3F = FH [24] & FC_2 = FH [19] (why_cong_resolution)⇒  FC_2 = FO_3 [182]
156. FC_2 = FO_3 [182] (r13)⇒  ∠FC_2O_3 = ∠C_2O_3F [183]
157. B,A,C_2 are collinear [18] & B,A,F are collinear [06] (i04)⇒  B,A,F,C_2 are collinear [184]
158. ∠FC_2O_3 = ∠C_2O_3F [183] & B,A,F,C_2 are collinear [184] (i08)⇒  ∠(AB-C_2O_3) = ∠C_2O_3F [185]
159. ∠(DF-C_2H) = ∠(C_2O_3-AB) [181] & ∠(C_2O_3-AB) = ∠FO_3C_2 [185] (why_eqangle_resolution)⇒  ∠FO_3C_2 = ∠(DF-HC_2) [186]
160. C_2O_3 ⟂ C_1O_3 [175] & ∠C_2HC_1 = ∠C_2O_3C_1 [150] (why_eqangle_resolution)⇒  ∠C_2HC_1 = ∠C_1O_3C_2 [187]
161. FC_1 = FH [17] (r13)⇒  ∠FC_1H = ∠C_1HF [188]
162. ∠FC_1H = ∠C_1HF [188] & B,A,F,C_1 are collinear [173] (i08)⇒  ∠(AB-C_1H) = ∠C_1HF [189]
163. ∠(AB-C_1H) = ∠C_2O_3H [152] & ∠(AB-C_1H) = ∠C_1HF [189] (why_eqangle_resolution)⇒  ∠C_1HF = ∠C_2O_3H [190]
164. ∠C_2HC_1 = ∠C_1O_3C_2 [187] & ∠C_1HF = ∠C_2O_3H [190] (r09)⇒  ∠C_2HF = ∠C_1O_3H [191]
165. ∠C_2HF = ∠C_1O_3H [191] & ∠(C_2O_3-DF) = ∠C_1O_3H [177] (why_eqangle_resolution)⇒  ∠(O_3C_2-DF) = ∠C_2HF [192]
166. ∠FO_3C_2 = ∠(DF-HC_2) [186] & ∠(O_3C_2-DF) = ∠C_2HF [192] (r09)⇒  ∠O_3FD = ∠DFH [193]
167. O_3F = HF [24] & ∠O_3FD = ∠DFH [193] (SAS 33)⇒  DO_3 = DH [194]
168. DA_1 = DH [09] & DO_3 = DH [194] & DA_2 = DH [11] (i01)⇒  A_2,H,O_3,A_1 are concyclic [195]
169. A_2,H,O_3,A_1 are concyclic [195] (r03)⇒  ∠A_2O_3H = ∠A_2A_1H [196]
170. ∠A_2O_3H = ∠A_2A_1H [196] & O_3,H,B are collinear [163] & B,A_1,A_2 are collinear [36] (i08)⇒  ∠(A_2O_3-BH) = ∠BA_1H [197]
171. ∠A_2O_3H = ∠A_2A_1H [196] & O_3,H,B are collinear [163] & B,A_1,A_2 are collinear [36] (i08)⇒  ∠(A_2O_3-A_1B) = ∠BHA_1 [198]
172. B,H,O_3 are collinear [163] & ∠(BH-A_2O_3) = ∠HA_1B [197] (why_eqangle6_resolution)⇒  ∠BO_3A_2 = ∠HA_1B [199]
173. A_1,A_2,B are collinear [36] & ∠(A_1B-A_2O_3) = ∠A_1HB [198] (why_eqangle6_resolution)⇒  ∠BA_2O_3 = ∠A_1HB [200]
174. ∠BO_3A_2 = ∠HA_1B [199] & ∠BA_2O_3 = ∠A_1HB [200] (Similar Triangles 35)⇒  O_3B:A_2B = A_1B:HB [201]
175. O_3C_2:O_3B = C_1H:AC_2 [172] & O_3C_2:C_2B = C_1H:HB [167] & O_3B:A_2B = A_1B:HB [201] (Ratio chase)⇒  AC_2:A_2B = A_1B:C_2B [202]
176. AC_2:A_2B = A_1B:C_2B [202] & C_1B = AC_2 [171] (why_eqratio6_resolution)⇒  BC_1:BA_2 = BA_1:BC_2 [203]
177. DE ∥ AB [84] & EF ∥ A_1B [47] (_why_eqangle_eqangle)⇒  ∠ABA_1 = ∠ABA_1 [204]
178. C_1,A,B are collinear [16] & A_1,A_2,B are collinear [36] & C_2,A,B are collinear [18] & ∠ABA_1 = ∠ABA_1 [204] (why_eqangle6_resolution)⇒  ∠C_1BA_2 = ∠C_2BA_1 [205]
179. BC_1:BA_2 = BA_1:BC_2 [203] & ∠C_1BA_2 = ∠C_2BA_1 [205] (Similar Triangles 39)⇒  ∠BC_1A_2 = ∠C_2A_1B [206]
180. ∠BC_1A_2 = ∠C_2A_1B [206] & B,A,C_1 are collinear [16] (i08)⇒  ∠(AB-A_2C_1) = ∠C_2A_1B [207]
181. C_1,B,A,C_2 are collinear [31] & A_1,B,A_2 are collinear [36] & ∠(AB-A_2C_1) = ∠C_2A_1B [207] (why_eqangle6_resolution)⇒  ∠C_2C_1A_2 = ∠C_2A_1A_2 [208]
182. ∠C_2C_1A_2 = ∠C_2A_1A_2 [208] (r04)⇒  C_1,C_2,A_1,A_2 are concyclic [209]
183. B,C,D are collinear [02] & B,A_1,C are collinear [08] (why_coll_resolution)⇒  B,A_1,D are collinear [210]
184. B,A_1,A_2 are collinear [36] & B,A_1,D are collinear [210] (why_coll_resolution)⇒  B,A_2,D are collinear [211]
185. B,A_2,C are collinear [10] & B,A_2,A_1 are collinear [36] & B,A_2,D are collinear [211] (i04)⇒  B,A_1,A_2,D are collinear [212]
186. B,A_1,A_2,D are collinear [212] (why_coll_resolution)⇒  D,A_2,A_1 are collinear [213]
187. D,A_2,A_1 are collinear [213] & DA_1 = DH [09] & DA_2 = DH [11] (why_midp_resolution)⇒  D is midpoint of A_2A_1 [214]
188. ∠(DF-HB) = ∠(HO_3-EA) [158] & A,B_1,E are collinear [94] (i08)⇒  ∠(DF-BH) = ∠(HO_3-AB_1) [215]
189. ∠(DF-BH) = ∠(HO_3-AB_1) [215] & A,B_1,C are collinear [12] & DF ∥ AB_1 [139] (i09)⇒  ∠(AC-BH) = ∠(HO_3-DF) [216]
190. EB_2 = EH [15] & EB_1 = EH [13] (why_cong_resolution)⇒  B_1E = B_2E [217]
191. B_1E = B_2E [217] & B_1O = B_2O [147] (r23)⇒  B_1B_2 ⟂ EO [218]
192. ∠(AC-BH) = ∠(HO_3-DF) [216] & AC ∥ DF [138] & HO_3 ∥ BH [160] (why_perp_resolution)⇒  DF ⟂ BH [219]
193. A,B_2,B_1 are collinear [30] & DF ∥ AB_1 [139] (why_para_resolution)⇒  B_1B_2 ∥ DF [220]
194. B_1B_2 ⟂ EO [218] & DF ⟂ BH [219] & B_1B_2 ∥ DF [220] (i00)⇒  EO ∥ BH [221]
195. BH ∥ HO_3 [160] & EO ∥ BH [221] ()⇒  HO_3 ∥ EO [222]
196. ∠(AC-BH) = ∠(HO_3-DF) [216] & AC ∥ DF [138] & BH ∥ EO [221] & HO_3 ∥ EO [222] (why_perp_resolution)⇒  DF ⟂ EO [223]
197. ∠(DE-HO_2) = ∠(HO_2-DE) [127] & HO_2 ∥ FO [86] (why_perp_resolution)⇒  DE ⟂ FO [224]
198. DF ⟂ EO [223] & DE ⟂ FO [224] (r43)⇒  FE ⟂ DO [225]
199. FE ⟂ DO [225] & EF ⟂ HO_1 [21] (i00)⇒  DO ∥ HO_1 [226]
200. DO ∥ HO_1 [226] & A,H,O_1 are collinear [51] (i05)⇒  DO ∥ AH [227]
201. ∠(AH-BC) = ∠(BC-AH) [38] & AH ∥ DO [227] & BC ∥ A_1B [39] (_why_perp_repr)⇒  DO ⟂ A_1B [228]
202. A_1,A_2,B are collinear [36] & DO ⟂ A_1B [228] (why_perp_resolution)⇒  OD ⟂ A_2A_1 [229]
203. D is midpoint of A_2A_1 [214] & OD ⟂ A_2A_1 [229] (r22)⇒  OA_1 = OA_2 [230]
204. C_1,C_2,A_1,A_2 are concyclic [209] & OC_1 = OC_2 [26] & OA_1 = OA_2 [230] (r50)⇒  OC_1 = OA_1 [231]
205. OC_1 = OB_2 [146] & OC_1 = OC_2 [26] & OC_1 = OA_1 [231] (i01)⇒  C_1,B_2,A_1,C_2 are concyclic [232]
206. C_1,B_2,A_1,C_2 are concyclic [232] & C_1,A_1,C_2,A_2 are concyclic [209] (i11)⇒  C_1,B_2,A_1,A_2 are concyclic [233]
207. C_1,B_2,A_1,C_2 are concyclic [232] & C_1,B_2,A_1,A_2 are concyclic [233] & C_1,B_2,C_2,B_1 are concyclic [145] (i11)⇒  A_2,B_2,C_1,A_1,B_1 are concyclic [234]
208. A_2,B_2,C_1,A_1,B_1 are concyclic [234] (why_cyclic_resolution)⇒  C_1,B_2,A_1,B_1 are concyclic [235]
209. C_1,B_2,A_1,C_2 are concyclic [232] & C_1,B_2,A_1,A_2 are concyclic [233] & C_1,B_2,A_1,B_1 are concyclic [235] (i11)⇒  C_1,C_2,B_1,B_2,A_1,A_2 are concyclic
==========================
\end{lstlisting}

\pagebreak
\section{\texttt{orthocenter\_aux Problem}}
\label{app:simple_problem}

Figures~\ref{fig:dependency_graph} and~\ref{fig:complex_symbols_graph} of the symbols graphs and the dependency graphs in the paper refer to the problem \texttt{orthocenter\_aux}, present in the \texttt{examples.txt} problem file already in the original AlphaGeometry codebase. Here we present a description of the problem.

The problem is to prove the existence of the orthocenter of the triangle, that is, the common intersection of the three heights:

\begin{problem} Given a triangle $ABC$ and $D$ the intersection of the heights of the triangle with respect to sides $AC$ and $AB$, prove that $D$ is also in the height relative to side $BC$. Take as an auxiliary point $E$, the foot of the vertex $B$ with respect to side $AC$.
\end{problem}

\paragraph{Explanation of AlphaGeometry's translation of the \texttt{orthocenter\_aux} problem:} Consider the triangle ABC and that
\begin{itemize}
    \item $D$ is the intersection of the line perpendicular to $AC$ through $B$ with the line perpendicular to $AB$ through $C$.
    \item $E$ is the intersection of line $AC$ with line $BD$.
\end{itemize}
Then, prove that line $AD$ is perpendicular to line $BC$.

\paragraph{Formal translation of the problem:} \

\begin{lstlisting}[language=jgex]
a b c = triangle a b c; 
d = on_tline d b a c, on_tline d c a b; 
e = on_line e a c, on_line e b d ? perp a d b c  
\end{lstlisting}

\pagebreak

\bibliographystyle{plainnat}
\bibliography{bibliography}

\end{document}